\documentclass[sn-mathphys-num]{sn-jnl}% Math and Physical Sciences Numbered Reference Style 
\usepackage{graphicx}%
\usepackage{multirow}%
\usepackage{amsmath,amssymb,amsfonts}%
\usepackage{amsthm}%
\usepackage{mathrsfs}%
\usepackage[title]{appendix}%
\usepackage{xcolor}%
\usepackage{textcomp}%
\usepackage{manyfoot}%
\usepackage{booktabs}%
\usepackage{algorithm}%
\usepackage{algorithmicx}%
\usepackage{algpseudocode}%
\usepackage{listings}%
\theoremstyle{thmstyleone}%
\theoremstyle{thmstyletwo}%

\theoremstyle{thmstylethree}%

\usepackage{pifont}% http://ctan.org/pkg/pifont
\usepackage{bbm}

\definecolor{myblue}{RGB}{55,126,184} % Customize the RGB values as needed
\definecolor{myorange}{RGB}{255,128,0} % Customize the RGB values as needed
\definecolor{mygreen}{RGB}{34,139,34} % Customize the RGB values as needed
\definecolor{myred}{RGB}{220,20,60} % Customize the RGB values as needed

\begin{document}

\title[ABIDES-Economist]{ABIDES-Economist: Agent-Based Simulator of Economic Systems with Learning Agents}


\author*[1]{\fnm{Kshama} \sur{Dwarakanath}}\email{kshama.dwarakanath@jpmorgan.com}

\author[2]{\fnm{Tucker} \sur{Balch}$^\dagger$}\email{tucker.balch@emory.edu}
% \equalcont{\footnotesize Tucker Balch's contributions to this paper were made while employed by JPMorgan Chase.}

\author[1]{\fnm{Svitlana} \sur{Vyetrenko}}\email{svitlana.vyetrenko@gmail.com}

\affil[1]{\orgdiv{AI Research}, \orgname{JP Morgan Chase}, \orgaddress{\city{San Francisco}, \state{California}, \country{USA}}}

\affil[2]{\orgdiv{Goizueta Business School}, \orgname{Emory University}, \orgaddress{\city{Atlanta}, \state{Georgia}, \country{USA.}}
\\{\footnotesize $\dagger$Tucker Balch's contributions to this paper were made while employed by JPMorgan Chase}}

% \affil[3]{\orgdiv{Department}, \orgname{Organization}, \orgaddress{\street{Street}, \city{City}, \postcode{610101}, \state{State}, \country{Country}}}


\abstract{
We present ABIDES-Economist, an agent-based simulator for economic systems that includes heterogeneous households, firms, a central bank, and a government. Agent behavior can be defined using domain-specific behavioral rules or learned through reinforcement learning by specifying their objectives. We integrate reinforcement learning capabilities for all agents using the OpenAI Gym environment framework for the multi-agent system. To enhance the realism of our model, we base agent parameters and action spaces on economic literature and real U.S. economic data.
To tackle the challenges of calibrating heterogeneous agent-based economic models, we conduct a comprehensive survey of stylized facts related to both microeconomic and macroeconomic time series data. We then validate ABIDES-Economist by demonstrating its ability to generate simulated data that aligns with the relevant stylized facts for the economic scenario under consideration, following the learning of all agent behaviors via reinforcement learning.
Specifically, we train our economic agents' policies under two broad configurations. The first configuration demonstrates that the learned economic agents produce system data consistent with macroeconomic and microeconomic stylized facts. The second configuration illustrates the utility of the validated simulation platform in designing regulatory policies for the central bank and government. These policies outperform standard rule-based approaches from the literature, which often overlook agent heterogeneity, shocks, and agent adaptability.
}




% \begin{enumerate}
%     \item \todo{Check references - address field for @book and @incollection}
%     \item \todo{Replace all agent parameters by $\alpha_{i,\mathrm{H}}$ or some other Greek letter instead of different letters for each parameter!}
%     \item \todo{Replace discount factor $\beta$ by $\gamma$}
% \end{enumerate}

\maketitle

\section{Introduction}\label{sec:intro}

Agent-based modeling (ABM) offers significant potential for advancing economics by defining agents and their interactions to produce complex emergent behaviors even from simple rules \cite{macal2005tutorial}. 
The main motivation behind applying ABMs to a domain is in the generation of emergent system-level behaviors that could not have been reasonably inferred from the underlying agent behaviors \cite{lavin2021simulation}.
ABMs have been applied in robotics \cite{vorotnikov2018multi}, financial markets \cite{byrd2019abides}, traffic management \cite{adler2005multi}, social networks \cite{gatti2014large}, and recently, in simulating social interactions with LLMs \cite{park2023generative}. %\cite{wang}
% Promoting agent-based modeling
Prominent economists highlight the benefits of ABMs in modeling complex scenarios accounting for human adaptation and learning \cite{farmer2009economy}. 
The field of Agent-Based Computational Economics advocates for ABMs' ability to simulate `turbulent' social conditions unseen in historical data, and to model dynamics out of equilibrium \cite{srbljinovic2003introduction,tesfatsion2006handbook}. 
\cite{hamill2015agent} promotes ABMs for bridging the gap between microeconomics (individual agent modeling) and macroeconomics (aggregate observations at system level).
\cite{arthur2021foundations} highlights the advantages of heterogeneous ABMs as a bottom-up approach towards modeling nuances of the real world more accurately. 

An agent is an entity that senses its environment to make a goal-oriented decision that is implemented by taking an action on the environment \cite{dorri2018multi}. Reinforcement learning (RL) deals with problems where an agent learns to act in an uncertain, dynamic environment through trial-and-error to maximize its objectives over a horizon \cite{kaelbling1996reinforcement}. When there are multiple agents that are attempting to learn to act in a common environment, they each introduce non-stationarity and (potential) partial observability for other agents \cite{busoniu2008comprehensive}. Multi-agent reinforcement learning (MARL) studies such problems by modeling the multi-agent system as a stochastic game where the state of the environment evolves in response to joint action across all agents \cite{littman1994markov,hu1998multiagent}. MARL is closely related to game theory which typically involves the study of multiple agents in static one-step or repeated tasks \cite{fudenberg1998theory,bowling2000analysis}. 

The field of industrial organization has a well-established literature on dynamic games, particularly focusing on the concept of Markov perfect equilibria within Markov Perfect models, which closely align with the Markov Decision Processes paradigm in RL \cite{pakes2001stochastic}. In MARL, the Markov assumption serves to restrict the space of game equilibria from which joint agent policies can be chosen, while allowing for unknown transition dynamics, unlike dynamic games. RL is a powerful tool for leveraging agent experiences from simulation runs to develop more flexible and robust agent behaviors. 
In economic systems, RL allows for flexible agent behaviors without restrictive parametric assumptions on how agents forecast the future. Specifically, the classical rational expectations assumption is an extreme version of such an assumption, where the forecasting mechanism is self-consistent: behavior based on these forecasts results in an environment that obeys these forecasted rules in expectation. Conversely, RL makes no such assumptions and allows for a more flexible (and more realistic \cite{moll2024trouble}) model of agent behavior, which can adapt more effectively to changes in policy and shock scenarios than traditional descriptive behavioral rules. This is particularly advantageous when agent objectives are easily formalized, as is often the case for firms and central banks.

Recent work has incorporated ABMs and RL in economic systems, but these are often restricted to simplistic 2D grid world environments \cite{zheng2022}, single agent (type) learning with rule-based background agents \cite{hill2021solving,chen2021deep,hinterlang2021optimal,brusatin2024simulating}, or a limited number of agent types learning alongside rule-based regulatory bodies \cite{curry2022analyzing,mi2023taxai}. Many of these approaches lack agent heterogeneity that is prevalent in real economies, and fail to model economic shocks within the system. 
Furthermore, there is no unified effort to define and apply calibration and validation practices for such ABMs with learning agents, as seen in traditional economic rule-based ABMs \cite{fagiolo2019validation}. 
Empirical calibration of simulation models is crucial for enhancing confidence in their results and their utility for economic decision-making \cite{werker2004empirical}. However, the high degree of freedom inherent in ABMs \cite{windrum2007empirical}, combined with the learning agents poses significant challenges to these efforts. 
Additionally, MARL can result in stability issues due to the interaction between agent decisions and the environment, alongside the explosion of parameters coming from ABMs, leading to an explosion in the space of possible equilibria that the system converges to. This is alongside learning instabilities inherent with MARL. We partially address these challenges in this work by introducing inductive biases to regularize the space of possible joint agent behaviors by assuming symmetry/exchangeability of policies per agent category \cite{pakes2001stochastic}, and via empirical validation of the resulting system with learned agents. These inductive biases serve as an equilibrium selection mechanism within the large space of joint agent policies considered by MARL within ABMs.

In this work, we introduce a customizable agent-based simulator for economic systems, implemented in Python, that incorporates multi-agent learning, agent heterogeneity, and economic shocks. We also examine stylized facts - recurring patterns observed in real economic data at both macro and micro levels — to guide the calibration of our ABM with learning agents. Once strategies are learned for all agents in our simulated economy, we validate these strategies against the stylized facts. Finally, we demonstrate the utility of our simulation platform and the learned policies of economic agents by benchmarking their performance against standard baselines. Our contributions are summarized as follows:
\begin{itemize}
\item \textbf{Development of a Versatile Agent-Based Simulator}: We have developed an agent-based simulator for economic systems that includes heterogeneous households, firms, a central bank, and a government. This simulator is highly versatile and customizable to various economic scenarios, allowing for the addition of regional agents, modification of inter-agent connections, and other system-specific configurations.

\item \textbf{Integration of Multi-Agent Reinforcement Learning}: In order to model agent adaptation to one another, we integrate reinforcement learning capabilities for all agents using OpenAI Gym-style environments within the multi-agent system. To efficiently scale multi-agent training, we adopt a shared policy network for all agents of a given type, incorporating their heterogeneity parameters as additional inputs. This along with communication of realistically shareable information between agents enable partial mitigation of non-stationarity and partial observability challenges associated with multi-agent learning. \footnote{We note that this also serves as a form of regularization that limits the space of possible agent strategies, and has previously been considered in the literature as the exchangeability assumption or symmetric Markov Perfect equilibrium in dynamic games \cite{pakes2001stochastic}.}

\item \textbf{Specification of Agent Heterogeneity and Action Spaces}: We specify agent heterogeneity parameters and action spaces grounded in economic literature and real-world economic data, to ensure realistic modeling.

\item \textbf{Survey and Validation of Stylized Facts}: We conduct a comprehensive survey of stylized facts (or empirical regularities) in historical economic data at both macroeconomic and microeconomic levels. This facilitates the validation of economic agent-based models. We validate our model by demonstrating its ability to generate simulated data that adheres to these facts, even as all agents in the platform are learning.

\item \textbf{Demonstration of Scalability and Configurability}: We demonstrate the simulator's scalability and configurability by simulating 10,503 microeconomic agents, including 250 farms, 1,250 companies, 250 retail stores, 8,750 households, one regional bank, a central bank, and a government, where each agent follows rule-based strategies. Additionally, we demonstrate scalability for multi-agent learning by simulating 112 macroeconomic agents, comprising 100 households, 10 firms, a central bank, and a government, each using deep reinforcement learning to arrive at objective-maximizing strategies.

\item \textbf{Utility for Monetary and Fiscal Policy Design}: We demonstrate the simulator's utility for monetary and fiscal policy design through two economic scenarios. In these scenarios, we compare the performance of learned central bank and government policies against standard rule-based policies, illustrating the superior performance of the learned policies.
\end{itemize}
% \begin{table}[tb]
%     \centering
%     \begin{tabular}{ccccc}\toprule
%         Simulator & Learning Agent Categories & Scale & Type of Calibration / Validation\\\midrule
%         ABIDES (and ABIDES-Gym) \cite{byrd2020abides,amrouni2021abides} & \\
%         (Py) MarketSim \cite{wah2013latency,mascioli2024financial} & \\
%         Phantom \cite{ardon2022phantom} &\\
%         \midrule
%         AI Economist \cite{zheng2022} & (Household, Planner) & (4, 1) & (intuitive trends; not-statistical properties) \\
%         Democratic AI \cite{koster2022human} & \\
%         MARL in RBC model \cite{curry2022analyzing} & (Household, Firm, Government) & (100, 10, 1) & \redcross\\
%         TaxAI \cite{mi2023taxai} & (Household, Government) & (10000, 1) & \greencheck (initial states, parameters)\\
%         Rational macro ABM \cite{brusatin2024simulating} & Firm & 20 & \redcross\\
%         Mortgage ABM \cite{ardon2024simulate,garg2024heterogeneous}\\
%         \midrule
%         ABIDES-Economist \cite{dwarakanath2024tax,dwarakanath2024empirical} & (Household, Firm, Bank, Government) & (,, 1, 1) & (parameters, stylized facts)\\\bottomrule
%     \end{tabular}
%     \caption{Comparison of our simulation platform against state-of-the-art in agent-based modeling for economics. Household is used synonymously with worker/consumer. }
%     \label{tab:my_label}
% \end{table}

\section{Literature Review}\label{sec:lit_review}
% We survey relevant past work in modeling economic systems using a class of macroeconomic models called Dynamic Stochastic General Equilibrium (DSGE) models \cite{an2007bayesian}. DSGE models are used in practice by Central Banks as tools for macroeconomic forecasting and policy analysis \cite{del2013dsge}. We also list literature using learning techniques in conjunction with economic modeling, before reviewing work in the domain of empirical game-theoretic analysis (EGTA) and its application in conjunction with reinforcement learning (RL). 

\subsection{Agent-Based Models (in Finance)}
Agent-Based models (ABMs) are powerful tools for simulating complex systems by defining individual agents and their interactions, even when governed by simple rules. These models often result in emergent behaviors that are far more intricate than the rules themselves \cite{macal2005tutorial}. ABMs have found applications across diverse domains, including robotics \cite{vorotnikov2018multi}, financial markets \cite{raberto2001agent,byrd2019abides}, traffic management \cite{adler2005multi}, social networks \cite{gatti2014large}.
Recent advances have leveraged large language models (LLMs) as agents to simulate realistic social interactions \cite{park2023generative} and achieve human-level performance in strategic games like Diplomacy by integrating natural language negotiation with planning and reinforcement learning \cite{meta2022human}.
In finance and economics, ABMs are particularly popular for analyzing and developing trading and investment strategies, where real-world testing can be prohibitively expensive or impractical. One of the most significant advantages of ABMs lies in their ability to capture emergent phenomena - outcomes driven by interactions among nonlinear and heterogeneous agents \cite{bonabeau2002agent}.

There is long-standing literature in agent-based finance following the Santa Fe Stock Market project \cite{palmer1994artificial,lebaron2000agent}. Past works consider agent-based financial modeling within continuous double auction (CDA) markets such as NASDAQ and the New York Stock Exchange \cite{byrd2020abides,vyetrenko2020get}, using the simulator to (1) investigate particular trading practices such as latency arbitrage \cite{wah2013latency}, spoofing \cite{wang2017spoofing}, market making \cite{wah2017welfare}; (2) to simulate rare events such as flash crashes \cite{paddrik2012agent,zhu2023once}; and (3) to design and test new financial policies before implementation in real markets \cite{dwarakanath2022equitable,dwarakanath2023transparency}. An alternate line of research models over-the-counter (OTC) markets such as foreign exchange markets where liquidity providers and takers interact in a decentralized manner with direct connections to each other (in absence of a centralized exchange) \cite{ardon2022phantom,vadori2024towards}. \cite{hugonnier2025economics} propose a search-theoretic framework to model OTC markets, documenting empirical regularities that are common across OTC markets. 
With access to these simulators, one can use reinforcement learning to learn approximately optimal strategies for different objectives, such as for market making \cite{spooner2018market,ganesh2019reinforcement,dwarakanath2021profit}, large order execution \cite{amrouni2021abides}, daily investment \cite{amrouni2021abides,mascioli2024financial}, as well as for mechanism design experiments \cite{darley2007nasdaq,dwarakanath2022equitable,dwarakanath2023transparency}.

\subsection{Economic Models}
Dynamic Stochastic General Equilibrium (DSGE) models are macroeconomic models widely used by Central Banks for macroeconomic forecasting and policy analysis \cite{del2013dsge}.
They are \textit{dynamic} as they model the evolution of economic observables over time, \textit{stochastic} in incorporating external random shocks to the economy. And, they model economies in \textit{general equilibrium} where the assumption is that supply equals demand for goods and labor. 
% , representing economic agents - households, firms, monetary, and fiscal authorities - at equilibrium.
Early work, like \cite{kydland1982time}, introduced a DSGE model with a representative household and firm, analyzing stochastic dynamics local to a balanced-growth path and perturbations around them. \cite{krusell1998income} overcomes the representative agent assumption of having a single household in the economy by incorporating household heterogeneity in income, wealth, and temporal preferences, using a continuum of households subject to employment shocks.

Modern macroeconomic modeling focuses on models estimated from real data \cite{christiano2005nominal,woodford2009convergence}. 
This is also seen in \cite{smets2007shocks}, where Bayesian techniques are used to estimate a DSGE model with a representative household and firm, incorporating price setting by firms and wage setting by labor unions. \cite{kaplan2018monetary} explored the impact of monetary policy on household consumption and labor, including dual-asset savings with transaction costs. 
Numerous software packages are available for estimating and solving DSGE models \cite{dynare:072,cao2023global}. 
The Federal Reserve Bank of New York provides public access to its DSGE model \cite{negro2015}, and forecasts \cite{frbny_dsge_jl}.

Despite the extensive literature, DSGE models rely on restrictive assumptions, such as representative agents, general equilibrium, and individual and aggregate rationality \cite{fagiolo2016macroeconomic}. The representative agent assumption is often employed because authors find that when focusing solely on aggregate macroeconomic behavior, a model with a representative agent performs comparably to more complex models. This finding hinges on the requirement of complete markets and the rational expectations assumption. However, these simplifying assumptions limit DSGE models' ability to capture the full complexity of real economies \cite{haldane2019drawing} and make them susceptible to model mis-specification errors \cite{farmer2009economy}. DSGE models have particularly struggled to simultaneously explain certain stylized facts observed in real economies, especially those related to heterogeneity \cite{fukac2006issues}. Nonetheless, these models are easier to calibrate than agent-based models due to their fewer parameters, restricting the space of equilibria for joint agent policies.

Numerous studies have examined expectation formation among economic agents, highlighting how individuals interpret the world and form expectations, which influences their economic activity. This underscores the limitations of rule-based behavioral modeling of agents by assigning parametric assumptions to agents' expectations, as restricting the complexity of their decisions. Critiques of the rational expectations model, where agents are assumed to have perfect forecasts of the world, have been noted \cite{evans2012learning, moll2024trouble}, advocating for the inclusion of adaptive expectations and learning for agents \cite{sargent1999conquest}.

\subsection{Agent-Based Models in Economics}
The field of Agent-based Computational Economics (ACE) employs ABMs to simulate interactions among economic agents, addressing key limitations of DSGE models including the inability to capture agent heterogeneity, adaptive behaviors, and out-of-equilibrium dynamics \cite{tesfatsion2006handbook}. ABMs provide a versatile framework to model complex, heterogeneous, and boundedly rational economic agents with diverse objectives \cite{stiglitz2018modern}. 
 Furthermore, \cite{moll2024trouble} critiques the assumption of rational expectations in heterogeneous macroeconomic modeling as unrealistic, suggesting that least-squares learning and reinforcement learning are promising approaches to overcome this challenge.

A prominent example of ABMs in economics is the EURACE project \cite{deissenberg2008eurace}, which aimed to build a large-scale model of the European economy. Early simulations explored labor market dynamics, involving capital and consumer goods firms alongside households employing rule-based strategies. Subsequent iterations extended the framework to encompass interactions between labor markets, industry evolution, credit markets, and consumption \cite{dawid2016heterogeneous}.

In studies of systemic risk and financial crises, ABMs have been used to model housing markets, which played a pivotal role in the 2007-2009 financial crisis \cite{geanakoplos2012getting,carro2023heterogeneous}. These models include mortgage-borrowing households (homeowners) with diverse characteristics who can choose to make payments, prepay loans, or face foreclosure, alongside mortgage lenders. Key scenarios involve examining how exogenous leverage or income shocks impact borrowers' loan repayment capacity \cite{geanakoplos2012getting,carro2023heterogeneous}.

In the context of climate economics, there has been recent work on using ABMs to simulate the impact of global climate negotiations and agreements on global temperatures. 
\cite{zhang2022ai} introduced a calibrated multi-region climate-economic-trade simulation platform to study negotiation protocols and their effect on fostering cooperation among global regions with differing economic policy objectives. Building upon traditional heterogeneous agent equilibrium models, such as \cite{nordhaus1996regional}, this platform incorporates negotiation protocols, and accounts for international trade and tariffs, offering a testbed for climate-economic policy design.

ABMs are increasingly being explored as experimental platforms for macroeconomic policy design across a range of applications, including fiscal and monetary policy, bank regulation, structural reforms in the labor market, and climate change policies. However, important challenges remain, particularly regarding their empirical validation, over-parameterization, estimation, and calibration \cite{fagiolo2016macroeconomic}.
Calibrating ABMs with numerous agents and diverse agent types using real data presents challenges due to the large degree of freedom arising from the variables and parameters that characterize agents' decision rules \cite{windrum2007empirical}. 

The most common method for validating the assumptions and effectiveness of an ABM as a realistic model of the true economic system is through indirect calibration. This involves assessing whether the model can replicate a set of statistical regularities observed in real economic data, known as \textit{stylized facts}, which are selected by the model designer for validation purposes.
Substantial work has focused on ABMs of endogenous growth and business cycles, which are empirically validated by replicating sets of microeconomic and macroeconomic stylized facts \cite{dosi2006evolutionary,dosi2017micro}. These models typically employ rule-based agents with predefined behaviors, providing a framework for comparing behavioral rules, particularly in the context of monetary and fiscal policies \cite{dosi2015fiscal}. For a comprehensive survey of ABMs in macroeconomic analysis, see \cite{dawid2018agent}.

%\cite{lan2022warpdrive} is an open source framework for fast deep multi-agent RL.

\subsection{Agent-Based Economic Modeling and Reinforcement Learning}
Since economic models often depict households as entities maximizing their discounted utility over time, reinforcement learning (RL) techniques are particularly well-suited for modeling household behavior \cite{atashbar2023ai}. For instance, \cite{chen2021deep} employed RL to derive consumption, saving, and labor strategies for a representative household in a DSGE model proposed in \cite{evans2005policy}. Similarly, \cite{hill2021solving} applied RL to learn consumption and labor strategies for heterogeneous households in macroeconomic models that incorporate epidemiological dynamics under equilibrium.
In the domain of housing markets, \cite{garg2024heterogeneous} utilized RL to develop policies for mortgage-borrowing households within an ABM encompassing heterogeneous households, mortgage servicers, mortgage owners, and the broader economy. Their experiments examined the impacts of exogenous income shocks on borrowers across income quartiles, revealing that lower-income borrowers were disproportionately affected. 

Beyond household strategies, RL has been used to optimize other economic strategies. For example, \cite{koster2022human} explored the design of a human-preferred revenue redistribution mechanism, while \cite{chen2021deep,hinterlang2021optimal} investigated central bank monetary policy, and \cite{zheng2022} examined government tax policy. 
\cite{koster2022human} addressed the problem of `value alignment', focusing on developing AI systems whose decisions align with human preferences. They used RL to design a shared revenue redistribution mechanism in an online investment game played by groups of humans. Their findings indicated that the RL mechanism, trained to maximize votes from virtual human players (modeled using sample human data), was preferred by real players over baseline mechanisms, including egalitarian and libertarian approaches.
\cite{hinterlang2021optimal} demonstrated that their RL-derived monetary policy outperformed traditional rule-based interest rate policies, like those in \cite{taylor1993discretion,nikolsko2021policy}, in achieving inflation and productivity targets \cite{svensson2020monetary}. Here, the environment (containing everything other than the central bank), modeled as a neural network fitted to historical U.S. data, predicted inflation and productivity in response to actions of the central bank.

Despite significant advancements, many RL studies focus on strategies for a single agent type or a small subset of agent types. However, a significant critique of modeling and learning economic agents in isolation comes from \cite{lucas1976econometric}, which highlights the inability of such models to account for how other agents react to changes in an agent’s policy. 
\cite{zheng2022} pioneered the use of multi-agent RL (MARL) in economic ABMs by exploring tax policy design with four agents and a planner. The planner optimized marginal tax rates to balance equality and productivity, while agents maximized utility based on their endowments.
\cite{curry2022analyzing} introduced a macroeconomic real-business-cycle ABM that applied MARL to 100 consumers, 10 firms, and a government. However, their model omitted unemployment modeling, the central bank’s role in monetary policy, used uniform tax redistribution, and lacked economic shocks. 
\cite{mi2023taxai} developed an ABM for taxation study, using MARL for 10,000 household agents that optimized consumption utilities while the government aimed to enhance social welfare and economic growth. Similarly, \cite{brusatin2024simulating} utilized MARL in a macroeconomic ABM with capital and credit markets to learn price-quantity strategies for 20 consumer-goods firms, while households, capital-goods firms, and banks followed fixed strategies.

While MARL applications in economic ABMs remain relatively nascent, this field is expanding \cite{lan2022warpdrive,dwarakanath2024abides,dwarakanath2024tax,dwarakanath2024empirical}. In this work, we develop an economic model with heterogeneous households, heterogeneous firms, a central bank, and a government. All agents learn and adapt using MARL, in presence of economic shocks. To reduce the space of possible joint agent policies, we assume symmetric/exchangeable policies across agents within each category.
Crucially, to calibrate the multi-agent economic system, we ground agent parameters and actions in real economic data. Subsequently, we validate the model where agents follow their learned policies, against a broad set of economic stylized facts which are also surveyed in this work.

\section{Multi-Agent Economic System}\label{sec:maes}
\begin{figure}[tb]
    \centering
    \includegraphics[width=0.9\linewidth]{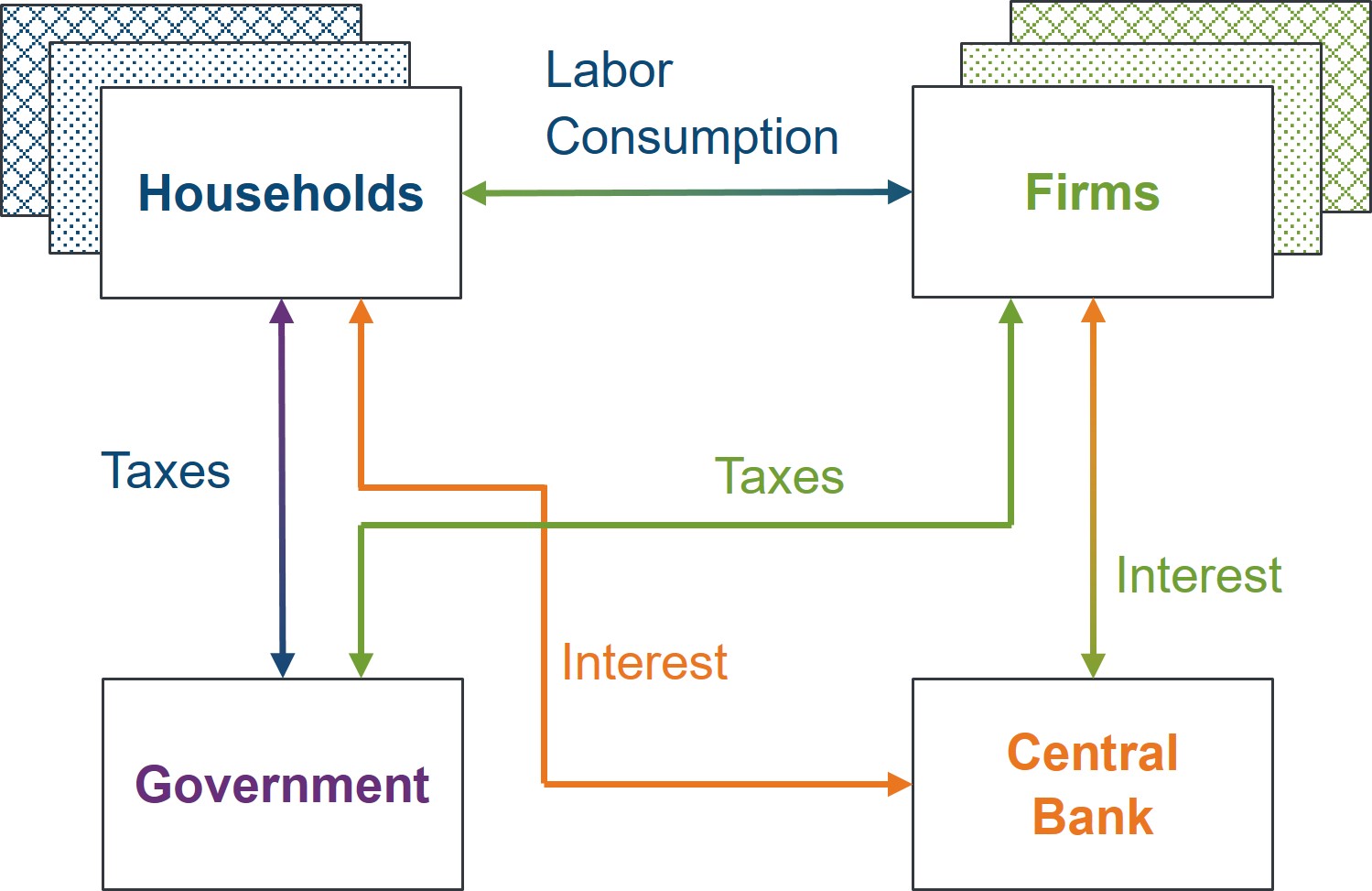}
    \caption{Agent types and interactions in ABIDES-Economist.}
    \label{fig:economy}
\end{figure}

Our economic model consists of four types of agents as shown in Figure \ref{fig:economy}.\begin{itemize}
    \item \textbf{Households} who are the consumers of goods and provide labor for the production of goods
    \item \textbf{Firms} who utilize labor to produce goods and pay wages
    \item \textbf{Central Bank} that monitors price inflation and production to set interest rate for household savings and firm deposits
    \item \textbf{Government} that collects income taxes from households and corporate taxes from firms, part of which could be redistributed as tax credits 
\end{itemize}
Each agent type has specific objectives and uses available economic information to adjust their actions to meet those objectives. Households aim to maximize utility from consumption and savings, adjusting their consumption based on prices, wages, interest rates, and tax information. Firms seek to maximize profits by setting prices and wages, informed by labor availability, past consumption, interest rates, and tax rates. The central bank targets inflation control and GDP growth, adjusting interest rates based on current inflation and GDP data. The government focuses on social welfare, setting tax rates and distributing tax credits based on household inequality and tax revenue.

Overall, each agent uses available economic information to adjust their actions. This behavior can be defined using functional rules that map input information to output actions, such as the Taylor rule for monetary policy or firm pricing using a fixed markup over production costs. While such predefined behavioral rules allow for faster simulations, they can limit agents' ability to achieve higher utility in shock scenarios, thereby restricting the robustness of their behaviors (as we demonstrate in our experiments).
In contrast, the availability of a simulator combined with advances in deep reinforcement learning (RL) enables us to learn these behavioral rules from simulated agent interaction data, including data from shock scenarios. Thus, while our simulator accommodates both rule-based and learning agents, we focus on agents that learn behaviors through interactions, using a multi-agent reinforcement learning (MARL) framework. 
In this MARL framework, each agent's long-term objectives are represented through a per-step reward function that is accumulated over time. And, we aim to learn policy functions (behavioral rules) that map observations to actions.

Formally, our economic model with multiple RL agents is represented as a Markov Game, where each agent has partial observability of the global system state \cite{littman1994markov,hu1998multiagent}. The global system state encompasses all information relevant to all agents in the system (e.g. savings of all households, inventories of all firms, etc.), as well as system-specific information like shock variables. 
So, the global system state along with actions of all agents fully determine the next global state, capturing the Markov assumption.
As expected, each agent only has access to partial information about the global state, including their own observables and publicly available data such as tax and interest rates.

A finite horizon Partially Observable Markov Game (POMG) is denoted by $\Gamma=\langle \mathcal{N},\mathcal{S},\lbrace\mathcal{A}_i\rbrace_{i=1}^n,\lbrace\mathcal{O}_i\rbrace_{i=1}^n,\mathbb{T},\lbrace \mathbb{O}_i\rbrace_{i=1}^n,\lbrace R_i\rbrace_{i=1}^n,\lbrace\beta_i\rbrace_{i=1}^n,H\rangle$ where \begin{itemize}
    \item $\mathcal{N}=\lbrace1,2,\cdots,n\rbrace$ is the set of agents
    \item $\mathcal{S}$ is the state space
    \item $\mathcal{A}_i$ is the action space of agent $i$ with $\mathcal{A}=\mathcal{A}_1\times\mathcal{A}_2\times\cdots\times\mathcal{A}_n$ denoting the joint action space
    \item $\mathcal{O}_i$ is the observation space of agent $i$ %for $i\in\mathcal{N}$ 
    \item $\mathbb{T}:\mathcal{S}\times\mathcal{A}\rightarrow\mathbb{P}\left(\mathcal{S}\right)$ is the transition function mapping the current state and joint action to a probability distribution over the next state
    \item $\mathbb{O}_i:\mathcal{S}\rightarrow\mathbb{P}\left(\mathcal{O}_i\right)$ is the observation function mapping the current state to a probability distribution over observations of agent $i$ %for $i\in\mathcal{N}$
    \item $R_i:\mathcal{S}\times\mathcal{A}\rightarrow\mathbb{R}$ is the reward function of agent $i$ %for $i\in\mathcal{N}$
    \item $\beta_i\in[0,1)$ is the discount factor of agent $i$\footnote{The symbol $\beta$ is used for the discount factor instead of the standard $\gamma$, as $\gamma$ is designated for a parameter related to households.}
    \item $H$ is the horizon
\end{itemize}
The objective of each agent $i\in\mathcal{N}$ in a POMG is to find a sequence of their own actions that maximizes their expected sum of discounted rewards over the horizon\begin{align}
    \max_{\left(a_i(0),\cdots,a_i(H-1)\right)}\ \mathbb{E}\left[\sum_{t=0}^{H-1}\beta_i^tR_i\left(s(t),a_1(t),\cdots,a_n(t)\right)\right]\nonumber
\end{align}
where $s(t+1)\sim\mathbb{T}\left(s(t),a_1(t),\cdots,a_n(t)\right)\ \forall t$. Here, $t$ represents a time step in the simulation, typically one quarter of a year in macroeconomic models.

Hence, an agent $i$ with a predefined behavioral rule has a fixed function $f$ that outputs its action given its observation as $a_i(t)=f(o_i(t))$ where $o_i(t)\in\mathcal{O}_i$ and $a_i(t)\in\mathcal{A}_i$. In a learning context, this function $f$ is updated over training steps using simulated agent interaction data. Without loss of generality, we describe the learning agent setup where we use index $i$ for households and $j$ for firms as we detail agent observations, actions, and rewards, thereby formalizing the POMG for our system.

\subsection{Households}\label{subsec:H}
% \cite{curry2022analyzing,hill2021solving,krusell1998income,evans2005policy}

Households are the consumer-workers in the economic system that provide labor for production at firms, while also consuming some of the produced goods. When employed, they are paid wages for their labor at their employer firm, and pay for the price of consumed goods. The government collects income taxes on their labor income, part of which could be redistributed back to households as tax credits in the subsequent time step. They also earn (accrue) interest on their savings (debt) from the central bank. These monetary inflows and outflows govern the dynamics of household savings/deposits from one time step to the next. 

The \textbf{observations} of household $i$ at time $t$ include tax credit $\kappa_{t,i}$, tax rate $\tau_{t,\mathrm{H}}$, interest rate $r_t$, prices of goods of all firms $\lbrace p_{t,j}:\forall j\rbrace$, wage at employer firm $\sum_jw_{t,j}e_{t,ij}$ where $e_{t,ij}$ is the employment indicator of household $i$ at firm $j$, 
% wages of all firms $\lbrace w_{t,j}:\forall j\rbrace$, their employment indicator at each firm $e_{t,ij}$ 
their monetary savings $m_{t,i}$ and their skills at all firms $\lbrace\omega_{ij}:\forall j\rbrace$. The employment indicator $e_{t,ij}=\begin{cases}
    1,\textnormal{ if household }i\textnormal{ is employed by firm }j\textnormal{ at time }t\\
    0,\textnormal{ otherwise}
\end{cases}$ captures if and where household $i$ is employed at time $t$. Note that $\sum_je_{t,ij}\leq1$ so that every household can be employed by at most one firm at time $t$. Household skills are used by firms in their hiring and firing decisions as described in section \ref{subsec:F}. When employed, each household provides $\bar{n}$ hours of labor at their employer firm.

The \textbf{actions} of household $i$ include the units of good requested for consumption at all firms $\lbrace c^{\textnormal{req}}_{t,ij}:\forall j\rbrace$. 

The \textbf{dynamics} related to household $i$ are given by\begin{align}
c_{t,ij}&=\min\left\lbrace c_{t,ij}^{\textnormal{req}},\left(Y_{t,j}+y_{t,j}\right)\cdot\frac{c_{t,ij}^{\textnormal{req}}}{\sum_kc_{t,kj}^{\textnormal{req}}}\right\rbrace\label{eq:H_dyn1}\\
    m_{t+1,i}&=(1+r_t)m_{t,i}+\sum_j\left(\bar{n}e_{t,ij}w_{t,j}-c_{t,ij}p_{t,j}\right)\nonumber\\
    &\quad-\tau_{t,\mathrm{H}}\cdot\sum_j\bar{n}e_{t,ij}w_{t,j}+\kappa_{t,i}\label{eq:H_dyn2}
\end{align}
where (\ref{eq:H_dyn1}) handles the case when the requested consumption per firm $j$ exceeds its inventory along with produced goods. Here, goods are distributed proportionally to households based on their requests, to give the realized consumption for household $i$ of goods of firm $j$ at $t$ as $c_{t,ij}$. (\ref{eq:H_dyn2}) is the evolution of savings from $t$ to $t+1$ based on interest earned/accrued, labor income, consumption spending, taxes paid on labor income, and tax credits received from the government. 
% The inflow for households comprises interest on savings, income from skilled labor at firms and any tax credits given by the government. Their outflow includes cost of consumption of goods and taxes paid on received income. 

The \textbf{reward} for household $i$ at $t$ is given by $u\left(\sum_jc_{t,ij},\bar{n}\sum_je_{t,ij},m_{t+1,i};\gamma_i,\nu_i,\mu_i\right)$ where \begin{align}
    u(c,n,m;\gamma,\nu,\mu)&=\frac{c^{1-\gamma}}{1-\gamma}-\nu n^2+\mu\cdot\mathrm{sign}(m)\frac{|m|^{1-\gamma}}{1-\gamma}\nonumber
\end{align}
with an isoelastic utility from consumption and savings, and a quadratic disutility of labor\footnote{Although we consider utility that is additive in consumption, labor and savings, our framework is flexible to use of any other. 
% \textcolor{red}{add citation} that could prevent the use of more classical DSGE models.
} \cite{evans2005policy}. Households can exhibit heterogeneity in their skill levels across firms and in the parameters of their utility functions. Table \ref{tab:heterogeneity} provides an overview of agent parameters, including those associated with heterogeneity.

% Parameters $\omega_{ij}$, $\gamma_i$, $\nu_i$ and $\mu_i$ control the heterogeneity of Household $i$.
\begin{table}[tb]
    \centering
    \begin{tabular}{ll}\toprule
        Agent & Parameter\\\midrule
        Household $i$ & $\omega_{ij}$: Skill level at firm $j$ \\
        & $\gamma_i$: Isoelasticity parameter\\
        & $\nu_i$: Weighting of labor disutility\\
        & $\mu_i$: Weighting of savings utility\\
        & $\beta_{i,\mathrm{H}}$: Discount factor\\
        & $\bar{n}$: Number of hours of labor at employer\\
        \midrule
        Firm $j$ 
        % & $f(c_{t-1},\cdots,c_{0})$: Demand forecasting function\\
        % & $h$: Demand forecasting horizon \\
        & $\omega_{\min}$: Minimum household skill to be hired by this firm\\
        & $\alpha_j$: Production elasticity for labor\\
        & $\rho_j,\bar{\varepsilon}_{j},\sigma_{j}$: Parameters of the exogenous shock process\\
        & $\chi_j$: Weighting of inventory risk\\
        % & $\varrho_j$: Weighting of savings utility\\
        & $\beta_{j,\mathrm{F}}$: Discount factor\\
        \midrule
        Central Bank 
        % & $g(u_{t-1},\cdots,u_{t-k})$ : Long-run unemployment rate\\
        % & $k$ : Horizon to compute $u_t^\star$ \\
        & $\pi^\star$: Target inflation \\
        & $\lambda$: Weighting factor for production \\
        & $\beta_{\mathrm{CB}}$: Discount factor\\
        \midrule
        Government & $l_{t,i}$: Weighting of households for social welfare\\
        & $\xi$: Portion of collected tax redistributed among households \\
        & $\theta$: Weighting of household utility in government reward\\
        & $\beta_{\mathrm{G}}$: Discount factor\\
        \bottomrule
    \end{tabular}
    \caption{Parameters for agents in our economic model. }
    \label{tab:heterogeneity}
\end{table}

\subsection{Firms}\label{subsec:F}
% \cite{curry2022analyzing,hill2021solving}
Firms are the producer-employers in the economic system that use household labor to produce goods for consumption.
They forecast consumption demand to set desired production and employment levels based on which they hire/fire households based on their skills. Labor from employed households is used to produce goods subject to an exogenous, stochastic production factor that captures any external shocks \cite{hill2021solving}. Firms accumulate inventory when they produce more goods than consumed by households, which they seek to minimize. They profit on revenue from prices paid by households for consumed goods, and pay wages for labor received from employed households. They pay taxes to the government on non-negative profits, and earn (accrue) interest on their deposits (debt) from the central bank. These monetary inflows and outflows govern the dynamics of firm deposits from one time step to the next. 

The \textbf{observations} of firm $j$ at time $t$ include tax rate $\tau_{t,\mathrm{F}}$, interest rate $r_t$, total household labor $\bar{n}\sum_ie_{t,ij}$, total consumption $\sum_ic_{t,ij}$, exogenous shock $\varepsilon_{t,j}$, exogenous production factor $\epsilon_{t-1,j}$, previous wage $w_{t,j}$, previous price $p_{t,j}$, inventory $Y_{t,j}$, and deposits $d_{t,j}$.

The \textbf{actions} of firm $j$ include wage per unit of labor $w_{t+1,j}$ and price per unit of good $p_{t+1,j}$ that go into effect at the next time step. 

The \textbf{dynamics} of quantities related to firm $j$ are given by\begin{align}
    \hat{C}_{t,j}&=f(C_{t-1,j},\cdots,C_{0,j})\label{eq:F_dyn1}\\
\hat{y}_{t,j}&=\max\lbrace0,\hat{C}_{t,j}-Y_{t,j}\rbrace\label{eq:F_dyn2}\\
\hat{N}_{t,j}&=\hat{y}_{t,j}^{\frac{1}{\alpha_j}}\label{eq:F_dyn3}\\
\epsilon_{t,j}&=\left(\epsilon_{t-1,j}\right)^{\rho_j}\exp\left(\varepsilon_{t,j}\right)\label{eq:F_dyn4}\\
y_{t,j}&=\epsilon_{t,j}N_{t,j}^{\alpha_j}\textnormal{ where }N_{t,j}=\bar{n}\sum_ie_{t,ij}\label{eq:F_dyn5}\\
Y_{t+1,j}&=Y_{t,j}+y_{t,j}-\sum_ic_{t,ij}\label{eq:F_dyn6}\\
d_{t+1,j}&=(1+r_t)d_{t,j}+p_{t,j}\sum_{i}c_{t,ij}-w_{t,j}\sum_i\bar{n}e_{t,ij}\nonumber\\
&\quad-\tau_{t,\mathrm{F}}\cdot\max\lbrace0,p_{t,j}\sum_{i}c_{t,ij}-w_{t,j}\sum_i\bar{n}e_{t,ij}\rbrace\label{eq:F_dyn7}
\end{align}
At the beginning of time step $t$, each firm $j$ forecasts net consumption demand from all households $\hat{C}_{t,j}$ based on previous values as in (\ref{eq:F_dyn1}), to compute the desired production $\hat{y}_{t,j}$ based on current inventory in (\ref{eq:F_dyn2}) and desired labor hours from employees $\hat{N}_{t,j}$ in (\ref{eq:F_dyn3}). The hiring/firing process for firms works as follows. If the firm requires more labor than can be provided by current employees $\hat{N}_{t,j}>\bar{n}\sum_ie_{t-1,ij}$, it sends hiring requests to all unemployed households with high skill $\omega_{ij}\geq\omega_{\min}$\footnote{Although it is straightforward to assign different minimum skill thresholds across firms, we adopt a uniform threshold for simplicity.}. Each unemployed household chooses among all hiring firms to provide labor to that firm at which it has highest skill. If the firm requires less labor than will be provided by current employees $\hat{N}_{t,j}<\bar{n}\sum_ie_{t-1,ij}$, it sends firing requests to all extra employees with lowest skills, ensuring it has at least one employee. 

After hiring/firing/not changing employees, the firm produces with employee labor per a Cobb-Douglas production function with elasticity parameter $\alpha_j\in[0,1]$ \cite{cobb1928theory} as in (\ref{eq:F_dyn5}). Here, $\epsilon_{t,j}$ represents an exogenous production factor following a log-autoregressive process with coefficient $\rho_j\in[0,1]$ as in (\ref{eq:F_dyn4}), with $\epsilon_{0,j}=1$ and $\varepsilon_{t,j}\sim\mathcal{N}\left(\bar{\varepsilon}_{j},\sigma^2_{j}\right)$ being an exogenous shock.
% $\epsilon_{t,j}$ denotes the exogenous production factor that follows a log-autoregressive process (\ref{eq:F_dyn1}) with coefficient $\rho_j\in[0,1]$ where $\varepsilon_{t,j}\sim\mathcal{N}\left(\bar{\varepsilon}_{j},\sigma^2_{j}\right)$ is an exogenous shock. 
The firm updates its inventory for the next time step based on current inventory and the difference between supply and demand as in (\ref{eq:F_dyn6}). The deposits of the firm evolve from $t$ to $t+1$ based on interest earned/accrued, profits, and taxes paid to the government on non-negative profits as in (\ref{eq:F_dyn7}).

The \textbf{reward} for firm $j$ at $t$ is given by \begin{align}
    p_{t,j}\sum_{i}c_{t,ij}-w_{t,j}\sum_i\bar{n}e_{t,ij}-\chi_jp_{t,j}Y_{t+1,j}\nonumber 
    % +\todo{\varrho_jd_{t+1,j}}
\end{align}
where the first two terms represent monetary profits as the difference in revenue from consumed goods and wages paid, with the last term capturing the risk of accumulated inventory. 
Firms exhibit heterogeneity in their sectors, which is equivalently captured by the shock process and the production function that transforms labor into goods. The parameters related to firm heterogeneity are provided in Table \ref{tab:heterogeneity}.

% Parameters $\rho_{j}$, $\bar{\varepsilon}_{j,y}$, $\sigma_{j,y}$, $\alpha_j$ and $\chi_j$ control the heterogeneity of Firm $j$.

\subsection{Central Bank}
% \cite{hinterlang2021optimal,chen2021deep,svensson2020monetary}
The central bank is the regulatory agency that monitors the prices and production of goods to set interest rates for household and firm deposits. By changing the interest rate on deposits, it affects the consumption patterns of households, which in turn affect the prices of goods produced by firms. The central bank seeks to set interest rates to meet inflation targets and boost production. Although it is uncommon for standard economics papers to focus on learning a policy for the Central Bank, our simulator retains this capability for generality of application e.g. learning monetary policy in presence of rule-based household and firm agents. Previous studies have explored similar learning models for central banks, as seen in \cite{sargent1999conquest, evans2012learning, hinterlang2021optimal}.

The \textbf{observations} of the central bank at time $t$ include previous interest rate $r_t$, total price of goods over the last five quarters $\lbrace \sum_jp_{t-k,j}:\forall k\in\{0,1,2,3,4\}\rbrace$, 
% number of employed households $\sum_i\sum_{j}e_{t,ij}$
and total production across firms $\sum_jy_{t,j}$.
% target long-run unemployment rate $u_t^\star=g(u_{t-1},\cdots,u_{t-k})$.

The \textbf{action} of the central bank includes the interest rate $r_{t+1}$ that goes into effect at the next time step. 

The \textbf{dynamics} related to the central bank are given by\begin{align}
\pi_t&=\frac{\sum_jp_{t,j}}{\sum_jp_{t-4,j}}\nonumber
    % u_t&=1-\frac{\sum_{i,j}e_{t,ij}}{n_{\mathrm{H}}}\nonumber
\end{align}
where $\pi_t$ is the annual inflation in total price.
% and $u_t$ is the unemployment rate with $n_{\mathrm{H}}$ number of households in the system.

The \textbf{reward} for the central bank is given by \begin{align}
    -\left(\pi_t-\pi^\star\right)^2+\lambda\left(\sum_jy_{t,j}\right)^2\label{eq:CB_reward}
    % -\lambda\left(u_t-u_t^\star\right)^2\nonumber
\end{align}
where $\pi^\star$ is the target inflation rate. And, $\lambda>0$ weighs the production reward in relation to inflation targeting.
% And, $\lambda>0$ weighs unemployment targeting in relation to inflation targeting.

\subsection{Government}\label{subsec:gov}
The government is the regulatory agency that collects taxes from households on their labor income and from firms on their profits, in order to maintain infrastructure. It sets tax rates and can choose to distribute a portion of the collected taxes back to households as tax credits in order to improve social welfare.

The \textbf{observations} of the government at time $t$ include the previous tax rates $\tau_{t,\mathrm{H}}$, $\tau_{t,\mathrm{F}}$, previous tax collected $\lbrace\tau_{t,\mathrm{H}}\sum_je_{t,ij}\bar{n}w_{t,j}:\forall i\rbrace$, $\lbrace\tau_{t,\mathrm{F}}\max\lbrace0,p_{t,j}\sum_ic_{t,ij}-w_{t,j}\sum_i\bar{n}e_{t,ij}\rbrace :\forall j\rbrace$, previous tax credits $\lbrace\kappa_{t,i}:\forall i\rbrace$, and a time varying weight associated to each household in relation to social welfare $\lbrace l_{t,i}:\forall i\rbrace$. 
% \kd{We will elaborate more on the weight $l_{t,i}$ for household $i$ in later sections}.
Our framework allows the designer to choose weights $l_{t,i}$ based on their choice of social welfare metric e.g., $l_{t,i}\equiv1$ for the utilitarian social welfare function versus $l_{t,i}=\mathbbm{1}\lbrace i=\arg\min_km_{t,k}\rbrace$ for the Rawlsian social welfare function. 
% We choose $l_{t,i}$ to be a linear function of household savings at $t$ with parameters $\alpha_l>0,\beta_l>0$, and clipped to lie in the range $[l_1,l_2]$ with $l_2>l_1>0$ as \begin{align}
%     l_{t,i}&=\begin{cases}
%         \max\lbrace l_1,-\alpha_lm_{t,i}+\beta_l\rbrace,\textnormal{ if }m_{t,i}>0\\
%         \min\lbrace l_2,-2\alpha_lm_{t,i}+\beta_l\rbrace,\textnormal{ if }m_{t,i}\leq0\\
%     \end{cases}\label{eq:G_h_weight}
% \end{align}
% (\ref{eq:G_h_weight}) gives weights that decrease with an increase in household savings for when savings are positive. When savings are non-positive, the weight increases with increase in household debt. Thus, the government under-weighs households that have high savings and over-weighs households that have high debts while ensuring all households are weighted at least $l_1>0$, and no household gets weighted higher than $l_2$.

The \textbf{actions} of the government include the tax rates $\tau_{t+1,\mathrm{H}}$, $\tau_{t+1,\mathrm{F}}$, and the fraction of tax credit distributed to each household $i$, $f_{t+1,i}$ that go into effect at the next time step. 

The \textbf{dynamics} related to the government are given by\begin{align}
\kappa_{t+1,i}&=f_{t+1,i}\cdot\xi\cdot\left(\tau_{t,\mathrm{H}}\sum_{i,j} e_{t,ij}\bar{n}w_{t,j}+\tau_{t,\mathrm{F}}\sum_j\max\lbrace0,p_{t,j}\sum_ic_{t,ij}-w_{t,j}\sum_i\bar{n}e_{t,ij}\rbrace\right)\label{eq:gov_dyn}
\end{align}
where $f_{t,i}\in[0,1]$ with $\sum_if_{t,i}=1$ so that a portion $\xi\in[0,1]$ of all collected taxes are redistributed. (\ref{eq:gov_dyn}) gives the tax credit for household $i$ at $t+1$ as a fraction $f_{t+1,i}$ of the $\xi$ portion of total tax collected in step $t$.

% The \textbf{reward} for the government is a measure of household social welfare, computed herein as a weighted sum of tax credits distributed to households as \begin{align}
% \sum_il_{t,i}\kappa_{t,i,}\nonumber
% \end{align}
The \textbf{reward} for the government is a measure of household social welfare, computed herein as a weighted sum of household utilities and tax credits as \begin{align}
    \theta\sum_il_{t,i}R_{t,i,\mathrm{H}}+(1-\theta)\sum_il_{t,i}\kappa_{t,i}\label{eq:G_reward}
\end{align}
where $l_{t,i}$ is the weight associated to household $i$, $R_{t,i,\mathrm{H}}=u\left(\sum_jc_{t,ij},\bar{n}\sum_je_{t,ij},m_{t+1,i};\gamma_i,\nu_i,\mu_i\right)$ is the reward function measuring the utility for household $i$ at time $t$, and $\theta\in[0,1]$ weighs household utility relative to tax credits. 
% Therefore, we use the household reward as a measure of household utility that is weighted by the importance factor $l_{t,i}$ for household $i$ to capture social welfare in the reward for the government.

% \kd{Add note on why we don't look at incorporating equilibrium conditions: not restrict agent policies.}
% \subsection{Equilibrium Conditions}
% \begin{itemize}
%     \item \textbf{Goods Market Clearing} $y_{t,j}=\sum_ic_{t,ij}$
%     \item \textbf{}
% \end{itemize}

\section{ABIDES-Economist Simulator}\label{sec:abides_econ_sim}
Our simulator is based on ABIDES, an agent-based interactive discrete event simulator that has been widely used to simulate financial markets with different types of trading agents \cite{byrd2019abides}. Agents in ABIDES have access to their internal states, and can receive information about other agents via messages. A simulation kernel handles message passing between agents, and runs simulations over a specified time horizon while maintaining timestamps for all agents and the simulation itself. We now describe the key components of setting up and running a simulation in ABIDES-Economist.

\subsection{Agent Configuration}
ABIDES-Economist defines a distinct agent class for each agent type described in Section \ref{sec:maes}, initialized with default parameters obtained from the literature. Table \ref{tab:parameters} outlines the default parameter values and their sources used in our simulator.
For every simulation run, users must specify the simulation horizon in quarters, the number of agents within each type, and any agent heterogeneity parameters that deviate from the default values. When modeling agent heterogeneity or testing hypothetical economic scenarios, the default parameters are replaced by the specified values.
\begin{table*}[h!]
    \centering
    \begin{tabular}{llll}\toprule
        Agent & Parameter & Value & Source \\\midrule
        Household $i$ & $\omega_{ij}$ & i.i.d $\mathcal{N}(1.0,0.3)$ & \\
        & $\gamma_i$ & $0.33$ & \cite{chen2021deep}\\
        & $\nu_i$ & $0.50$ & \cite{chen2021deep}\\
        & $\mu_i$ & $0.10$ & \cite{chen2021deep}\\
        & $\beta_{i,\mathrm{H}}$ & $0.99$ & \cite{chen2021deep}\\
        & \multirow{2}{*}{$\bar{n}$} & \multirow{2}{*}{$480$} & 40 hours/week \\
        & & & $\approx$ 480 hours/quarter\\
        \midrule
        Firm $j$ & $\omega_{\min}$ & 1.0 & Mean of each $\omega_{ij}$\\
        & $\alpha_j$ & i.i.d $\mathcal{U}[0.05,1.0]$ & \cite{bls_ind_output}, see appendix \ref{sec:app:firm_alpha}\\
        & $\rho_j,\bar{\varepsilon}_{j},\sigma_{j}$ & $0.97,0.00,0.10$ & \cite{hill2021solving}\\
        & $\chi_j$ & $0.50$ & \\
        % & $\varrho_j$ & $0.0$ & \\
        & $\beta_{j,\mathrm{F}}$ & $0.99$ & \\
        & \multirow{ 2}{*}{$f(c_{t-1},\cdots,c_{0})$} & Exponential Moving Average & \multirow{ 2}{*}{\cite{dosi2006evolutionary}}\\
        & & with half-life of 4 quarters & \\
        \midrule
        Central Bank 
        % & $g(u_{t-1},\cdots,u_{t-k})$ & Moving average & \\
        % & & $k$ & 20 quarters (5 years) & \\
        & $\pi^\star$ & $1.02$ & \cite{svensson2020monetary,hinterlang2021optimal}\\
        & $\lambda$ & $1.00$ & \cite{hinterlang2021optimal}\\ %\cite{svensson2020monetary}
        & $\beta_{\mathrm{CB}}$ & $0.99$ & \cite{hinterlang2021optimal}\\
        \midrule
        Government & \multirow{ 2}{*}{$l_{t,i}$} & $l_{t,i}=\frac{1}{m_{t,i}+\max_i\lbrace-m_{t,i}\rbrace+\epsilon_l}$,& Inverse-income\\
        & & normalized by $\sum_kl_{t,k}$ with $\epsilon_l=1.0$ & weights\\ 
        & $\xi$ & $0.10$ & \cite{irs_refunds}$^\ast$\\
        & $\theta$ & $1.00$ & \\
        & $\beta_{\mathrm{G}}$ & $0.99$ & \\
        \bottomrule
    \end{tabular}
    \caption{Default agent parameters in ABIDES-Economist. \\{\footnotesize$^\ast$Based on IRS data from 2023, 15\% of the income taxes collected from businesses and individuals were refunded to individuals \cite{irs_refunds}. We assume that two-thirds of these refunds, equivalent to 10\%, are related to tax credits, while the remaining portion is due to excess tax refunds.}}
    \label{tab:parameters}
\end{table*}

% \subsection{Agent communication}
Agents in our simulator can only access their internal states, so any information from other agents must be explicitly requested through message-based communication. When a message request for information is sent, the recipient responds by sharing the relevant part of their internal state with the sender.
To configure agents, we establish message-based communication channels based on the granularity and dynamics of the economic system. For instance, a household agent sends a message to each firm agent requesting the price of its goods. The firm agent responds with this information, which the household incorporates into its observations. Conversely, households only communicate with their current employers to obtain information about wages.

This messaging scheme ensures that all features in an agent’s observation that are external to itself are dynamically acquired through inter-agent communication. 
The ABIDES kernel processes these messages sequentially, resulting in simulation run time complexity that is linear in the number of messages exchanged.
To optimize performance, we minimize the number of messages in the system. Specifically, regulatory bodies communicate with households and firms through one-way messages, and only employee households exchange labor-related information with their employers. This design reduces overhead while preserving the necessary information flow for accurate agent behavior and system dynamics.

\subsubsection{Temporal Progression in the Simulation}\label{subsubsec:temp_progression}
Here is how the economic simulation proceeds from one time step $t$ (think quarter of year) to the next over a specified time horizon. At the start of the simulation, \begin{itemize}
    \item Households start with \$0 savings, with i.i.d skills sampled at the beginning of every training episode as $\omega_{ij}\sim\mathcal{N}(1.0,0.3)$. 
    \item Firms start with 0 units of inventory and \$0 deposits, with production elasticity i.i.d sampled at the beginning of every training episode as $\alpha_j\sim\mathcal{U}[0.05,1.0]$. Also, they each draw independent samples of price and wage for $t=0$ from a uniform distribution over their action space.
    \item All households are distributed among firms for employment uniformly at random, so that every household is employed at $t=0$.
    \item Central Bank samples the initial interest rate for $t=0$ from a uniform distribution over its action space.
    \item Government sets default tax rate and gives out \$0 of tax credits for $t=0$.
\end{itemize}
At each time step $t\geq0$,
\begin{enumerate}
    \item Each firm computes expected demand for this step based on past demand to compute desired number of employees (\ref{eq:F_dyn1}) - (\ref{eq:F_dyn3}). 
    \item Each firm sends out employment decisions of hiring, firing or status quo based on skills of households.
    \item Unemployed households choose from hiring firms based on highest skill match.
    \item Each firm uses labor from employee households to produce goods (\ref{eq:F_dyn4}) - (\ref{eq:F_dyn5}), and pays wages to employees.
    \item Each household observes tax rate, tax credits, interest rate, prices, and wage to decide on requested consumption.
    \item Each firm fulfills consumption (\ref{eq:H_dyn1}), updates its inventory (\ref{eq:F_dyn6}), and pays taxes to the government (\ref{eq:F_dyn7}).
    \item Each household updates savings based on realized consumption (\ref{eq:H_dyn2}) and pays taxes to the government (\ref{eq:H_dyn2}).
    \item Each firm sets price, wage for the next step based on consumption, labor in this step. 
    \item Central Bank monitors firm prices until this step and productions at this step to set interest rate for next step.
    \item Government collects taxes to set tax rate and distribute credits for next step (\ref{eq:gov_dyn}).
\end{enumerate}

\subsection{Scaling of Simulation Run Time with Agent Count}
Simulation run time within ABIDES-Economist is influenced by three main factors: agent initialization, message processing, and variable/data structure operations.
\begin{itemize}
    \item \textbf{Agent Initialization}: This involves constructing objects for each agent in the configuration and scales linearly with the total number of agents.
    \item \textbf{Message Processing}: This accounts for the transmission of information between agents via messages. It is the most significant factor, accounting for over 80\% of the total simulation run time. Recall that any information external to an agent must be conveyed via a message object from the sender agent who owns that information. The ABIDES kernel manages message passing between agents by processing them sequentially, resulting in a run time contribution that is linear in the number of messages processed. The number of messages is determined by the specific simulation dynamics described in Section \ref{subsubsec:temp_progression}. The greatest message volume complexity arises from price communications between producers and consumers, which scales as $\mathcal{O}(mn)$, where $m$ is the number of firms and $n$ is the number of households. This, combined with labor/employment interactions, leads to quadratic scaling in run time with the total number of agents.
    \item \textbf{Data Structure Operations}: These deal with compute operations within agents, and have been optimized using more efficient alternatives. For instance, we employ custom implementations of the Exponential Moving Average instead of the pandas version and replace \verb|numpy.sum()| with Python’s built-in \verb|sum()| function for lists, among other optimizations.
\end{itemize}
In summary, the simulation run time per simulation time step within ABIDES-Economist scales quadratically with the total number of agents, and accumulates over the time steps in the simulation horizon.

To assess the scalability of our simulation platform, we simulate a regional economy with $m$ farms, $5m$ companies, $m$ retail stores, $35m$ households (five times as many households as employers), along with one regional bank, a central bank, and a government, where each agent follows rule-based strategies. 
Here, farms and companies fall within the firm category mentioned throughout this paper, with retail stores sourcing products from farms and selling to households, while companies sell directly. Farms, companies, and retail stores act as employers, with households serving as potential employees.
The regional agents including farms, companies, retail stores, households and regional bank update their actions every two weeks, while the federal agents including the central bank and the government update their actions every quarter. Hence, the regional agents act 26 times a year while the federal agents act four times a year.

The objective of this agent configuration is to test scalability in a scenario with increased inter-agent communication and to demonstrate the customizability of our simulator, which allows for adaptation of the agent configuration to fit specific scenarios of interest. The increase in inter-agent communication arises from two sources. Firstly, these regional agents act more frequently than the macroeconomic agents presented earlier, which act once per quarter. Secondly, there is an increase in inter-agent connections from introducing a retail firm layer between households and farms and designating the regional bank to handle the banking needs of all microeconomic agents, while the central bank focuses on macroeconomic monetary policy.

\begin{figure}[h!]
    \centering
    \includegraphics[width=0.9\linewidth]{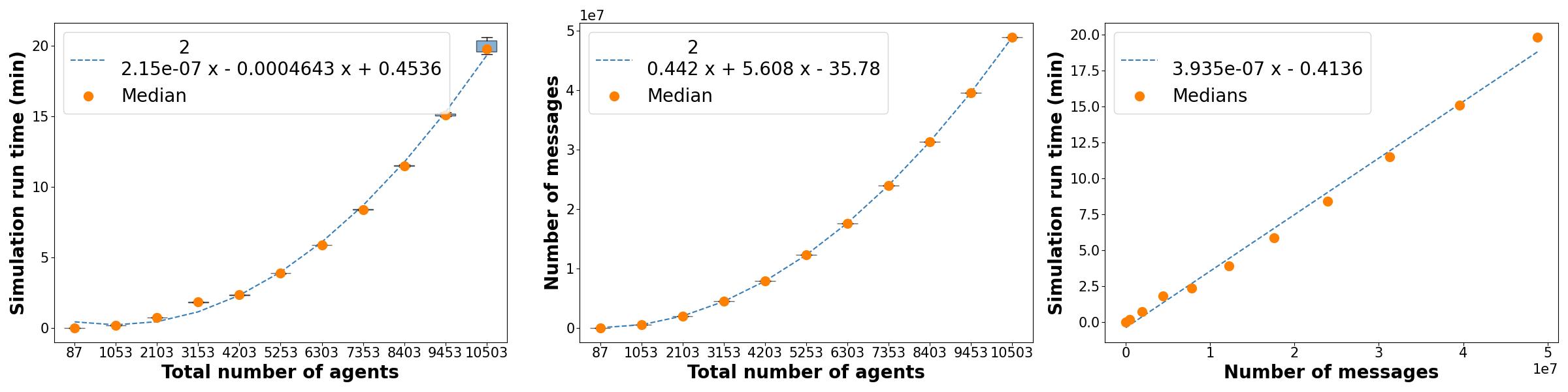}
    \caption{Simulation run time and number of messages processed as functions of agent count. Note that the run time scales quadratically with the agent count and linearly with the number of messages processed. }
    \label{fig:scalability}
\end{figure}

We vary $m$ in the range $\lbrace2,25,50,75,100,125,150,175,200,225,250\rbrace$ and plot the simulation run time and total number of messages processed per biweekly period across 10 instantiations of each configuration in Figure \ref{fig:scalability}. We used an AWS EC2 instance (type c5.4xlarge) with 16 vCPUs and 32GB RAM. The first subplot of Figure \ref{fig:scalability} shows boxplots of the simulation run time as a function of the total agent count, displaying a quadratic relationship. The second subplot shows boxplots of the number of messages processed by the kernel as a function of agent count, again displaying a quadratic relationship. The last subplot displays median simulation run time as a function of the median number of messages processed, showing a linear relationship. These trends align with our earlier explanation that sequential message processing is the primary factor influencing the scalability of simulations as the agent count increases.

Notably, we successfully simulated a model with up to 10,503 rule-based agents, exchanging nearly 50 million messages per simulation step of two weeks, with a run time of around 20 minutes on the specified instance. We capped the agent count at 10,000 because increasing it beyond 15,000 led to out-of-memory issues on the instance from instantiating the large number of agent objects.
Further reductions in run time can be achieved in economic scenarios with fewer inter-agent connections or where these connections are not continuously active. For instance, macroeconomic configurations with fewer inter-agent connections, fixed employment relationships without constant hiring and firing, or less frequent updates by regulatory bodies, such as the central bank and government, can improve efficiency. 
% As an example, the government may adjust tax rates annually, while households engage in consumption and work on a quarterly basis.

\subsection{Reinforcement Learning capabilities}

\subsubsection{Single-Agent Learning}
The original ABIDES framework was extended to incorporate a single reinforcement learning (RL) agent using an OpenAI Gym-style extension \cite{amrouni2021abides}. In this ABIDES-Gym setup, a Gym environment encapsulating the Markov Decision Process (MDP) for a single RL agent interacts with the core of ABIDES, which contains all rule-based agents, through a placeholder agent known as the Gym agent. The Gym agent serves as a proxy for the learning RL agent in its interactions with the rule-based environment.
In this setup, the Gym agent takes an action based on the current state of the world, prompting the remaining background agents to react and evolve the system to the next state. The Gym environment computes the corresponding next state and reward for the learning agent. Subsequently, any RL algorithm can be employed to derive an approximately optimal policy for the learning agent.

The learning process for a single RL agent involves three key steps: (1) MDP formulation by defining the states, actions, and rewards within the Gym environment; (2) capturing transition dynamics through the simulator's agent configuration, which includes all other background agents; and (3) selecting an appropriate RL algorithm to derive an optimal policy for the agent using environment interactions. The first two steps were discussed previously. 
For the third step, the choice of algorithm depends on the type of state and action space for the agent. Classical RL algorithms, like tabular Q-Learning \cite{watkins1992q}, are suited for discrete state and action spaces. In our scenarios, we typically encounter continuous states and/or actions, necessitating the use of neural network representations for the value function and/or policy, i.e., requiring deep RL algorithms.

Popular deep RL algorithms for continuous state/action spaces include deep Q-Learning \cite{mnih2015human}, vanilla policy gradient \cite{mnih2016asynchronous}, trust region policy optimization \cite{schulman2015trust}, and proximal policy optimization (PPO) \cite{ppo}. The latter three are policy gradient algorithms, which construct a parameterized representation of the policy using a neural network, with weights updated by gradient ascent on the value/advantage function \cite{mnih2016asynchronous}. These algorithms are favored for their learning stability, with PPO demonstrating reliably superior performance across benchmark RL tasks \cite{ppo}. 

% Most RL algorithms can be classified into three broad categories: (1) policy iteration methods, which alternate between estimating the value function under the current policy and improving the policy ; (2) policy gradient methods, which use an estimator of the gradient of the expected return (total reward) obtained from sample trajectories; and (3) derivative-free optimization methods, such as the cross-entropy method and covariance matrix adaptation, which treat the return as a black box function to be optimized in terms of the policy parameters \cite{schulman2015trust}. 

\subsubsection{Multi-Agent Learning}
We extend ABIDES-Gym to accommodate multiple RL agents by creating a multi-agent Gym environment that encapsulates a formalized Markov Game, with the Gym agent representing all learning agents in the system. This Gym agent interacts with the core of ABIDES which contains all rule-based agents, to generate the next system state. And, the Gym environment uses it to compute the learning agents' rewards. This setup allows for a mix of learning and rule-based agents. A similar extension to multi-agent reinforcement learning (MARL) within ABIDES has been applied in the financial domain \cite{dwarakanath2023transparency}.

The learning process for multiple agents involves three key steps: (1) Markov Game formulation by defining observations, actions, and rewards for all learning agents within the Gym environment; (2) capturing transition dynamics through the simulator's agent configuration, which includes all other rule-based agents; and (3) selecting an appropriate MARL algorithm to develop effective policies for multiple learning agents using environment interactions.
Here, the choice of algorithm is also influenced by the nature of agent interactions - collaborative (shared rewards), competitive (zero-sum rewards), or mixed (neither cooperative nor competitive), in addition to the observation and action spaces of the agents. In our scenario, agents have individual objectives that are neither cooperative nor competitive, placing their interactions in the mixed category. Given their continuous observation spaces, deep MARL algorithms are required.

From a theoretical perspective, as more agents are equipped with RL capabilities, the learning problem becomes more challenging due to non-stationarity. Transition dynamics and agent rewards are both functions of all agents' actions, which are adapted over time. As multiple agents simultaneously change their behaviors, each agent's environment (everything other than itself) becomes non-stationary, causing the optimal policy to change over time as other agents' policies evolve \cite{bowling2002multiagent}. This can result in learning process instabilities, preventing convergence.
Additionally, partial observability presents a challenge, as agents cannot see other agents' information and are unaware of the global system state. Their reward functions depend on the global state and other agents' actions, neither of which are observable. While algorithms exist for partially observable Markov Games, they require agents to maintain beliefs over the global state and other agents’ policies, which can become intractable in high-dimensional problems with many learning agents \cite{wong2023deep}.

In this work, we partially mitigate non-stationarity and partial observability challenges using the following techniques:
\begin{enumerate}
\item \textbf{Realistic Agent Communication}: Agents exchange realistically shareable information, which alleviates much of the observability limitations, especially concerning agent rewards. For example, firms share prices with consumers and wages with employees, while regulatory bodies share tax and interest rates with all agents.
\item \textbf{Partial Centralized Training}: We adopt a hybrid approach that falls in between fully centralized training and independent learning \cite{tan1993multi}. All agents of the same type share a common policy network, allowing them to exchange experience tuples with each other. This enables agents of the same type to share knowledge and diverse experiences, potentially accelerating the learning process \cite{wong2023deep}. These four shared policy networks corresponding to the four agent types, are then updated simultaneously using independent learning. 
As mentioned in the introduction \ref{sec:intro}, this inductive bias helps regularize the space of potential joint agent behaviors by assuming exchangeability of policies within each agent category \cite{pakes2001stochastic}. It also acts as an equilibrium selection mechanism within the extensive space of joint agent policies considered by MARL within ABMs.

\item \textbf{Learning Rates}: Through trial-and-error, we discovered that having firms and the central bank adapt more quickly than households and the government leads to improvement and approximate convergence of all agents' rewards over training episodes. We attribute this division into two tiers of learner groups - one adapting faster than the other - to the specific dynamics of our system, as described in Section \ref{sec:maes}. Notably, firms have greater authority in setting wages, with households having limited options to respond to low wages in the labor market, other than reducing consumption. These wages also influence the income taxes collected by the government and subsequent tax credits. Similarly, households lack alternative investment options beyond depositing their savings in the central bank, thus limiting their response to interest rates to varying consumption. Consequently, we hypothesize that having households and the government adapt more slowly than the more influential actors, such as firms and the central bank, stabilizes the learning process.
\end{enumerate}

\subsubsection{Scalability of Multi-Agent Learning}
We define training time as the total compute time required to achieve improvement and approximate convergence of rewards for all learning agents. Our goal is to quantify how training time scales as a function of (1) the total agent count and (2) the learning agent count. Training time can be decomposed as the product of two contributing factors: simulation time per training episode and the number of training episodes.
As discussed previously, simulation run time scales quadratically with the total number of agents. This relationship holds true even when multiple agents are learning, as it depends solely on the total number of agents in the configuration, irrespective of the number of learning agents.

Regarding the second contributing factor, training time scales linearly with the number of training episodes since episodes are executed sequentially. However, the number of episodes needed to reach a desired level of performance is heavily influenced by the specific dynamics and interactions among the agents, as well as the number of learning agents or even the total number of agents. For instance, cooperating agents who share information may require fewer training episodes when learning together compared to learning individually. Conversely, competing agents may need significantly more episodes when learning in the presence of other competing learners than when learning alone. In the next subsection, we conduct a small study of training time to provide further insight into this relationship within our economic system.

In summary, training time increases at least quadratically with the total number of agents when the number of learning agents is fixed. It also scales linearly with the number of training episodes required for approximate convergence. The relationship between the number of training episodes required and the total agent count or learning agent count is heavily dependent on the problem setup.
In practice, we utilize the RLlib package, which offers a wide selection of off-the-shelf single-agent and multi-agent RL algorithms \cite{rllib}. This package also provides scalability through parallel environment sampling, allowing the simultaneous collection of training batches that are used to update the policy networks. By sampling environments in parallel, we can significantly reduce the total training time compared to sequential sampling, by a factor proportional to the number of parallel environment samplers.

\subsubsection{Impact of Multiple Learning Agents}\label{subsubsec:impact_mal}
To better understand the impact of having multiple learning agents on training time and learned policies, we conduct a simple experiment. We consider three economic configurations, each comprising one household ($\omega=1$), one firm ($\alpha=\frac{2}{3}$ as in \cite{hill2021solving}, $\sigma=0.01$), one central bank, and one government ($\theta=0.2$), over a 10-year horizon.\begin{itemize}
    \item \textbf{Configuration 1 (LH)}: A policy is learned for the single household, while all other agents follow random policies. 
    \item \textbf{Configuration 2 (LF)}: A policy is learned for the single firm, with all other agents following random policies. 
    \item \textbf{Configuration 3 (LH + LF)}: Policies are learned for both the household and the firm, while the remaining agents follow random policies. 
\end{itemize}
The use of random policies for other agents simulates the presence of rule-based background agents, with the rule being defined by the random initialization of their policy neural networks.

\begin{figure}[tb]
    \centering
    \includegraphics[width=0.9\linewidth]{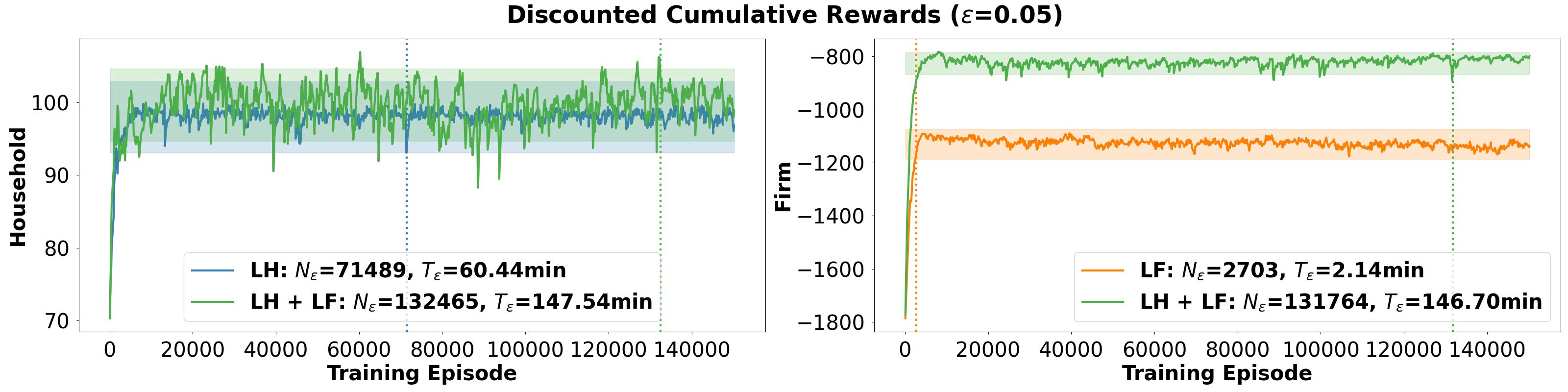}
    \caption{Training rewards for the \textbf{LH}, \textbf{LF}, and \textbf{LH + LF} configurations to demonstrate the impact of multiple learning agents. When both household and firm are learning, more training episodes are needed for the rewards to improve and converge as compared to when only one of them is learning.}
    \label{fig:mal_rewards}
\end{figure}

To assess the impact on training time, we examine the number of training episodes required for the rewards to improve and approximately converge. Specifically, after confirming reward improvement over training episodes, we determine the number of training episodes $N_\epsilon$ beyond which agents' rewards remain within an $\epsilon$-percentage range around the mean long-run reward. This provides an estimate of the required compute time $T_\epsilon$ needed to reach a certain level of training convergence in the different configurations. These experiments are run on an AWS EC2 instance (type c5.12xlarge) with 48 vCPUs and 96GB RAM, where we allocate 16 vCPUs per configuration. Figure \ref{fig:mal_rewards} shows the training rewards along with $N_\epsilon$ and $T_\epsilon$ for $\epsilon=5\%$.
When only the household is learning, it takes $N_\epsilon=71,489$ episodes (or 60.44 minutes of training time) to reach and remain within 5\% of the long-term rewards. Similarly, when only the firm is learning, it takes $N_\epsilon=2,703$ episodes (or 2.14 minutes of training time) to achieve approximate convergence. On the other hand, when both agents are learning, the number of episodes needed for convergence significantly increases to over 132,000 for the household reward and over 131,000 for the firm reward. This demonstrates that as the learning agent count increases even with the same total agent count, the training time for our economic system also increases. 
\begin{figure}[tb]
    \centering
    \includegraphics[width=0.45\linewidth]{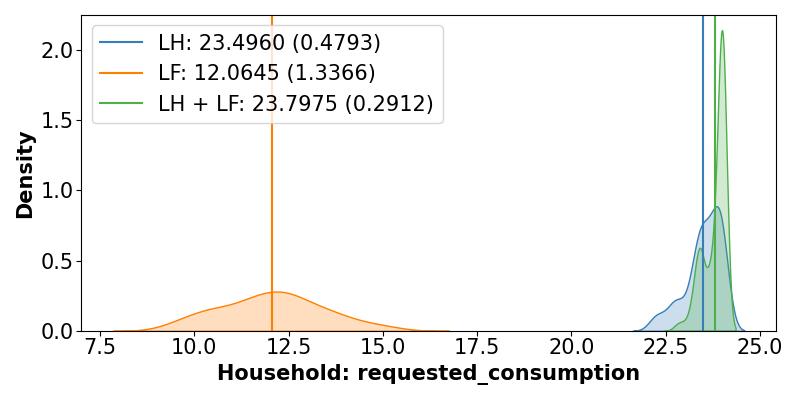}%
    \includegraphics[width=0.45\linewidth]{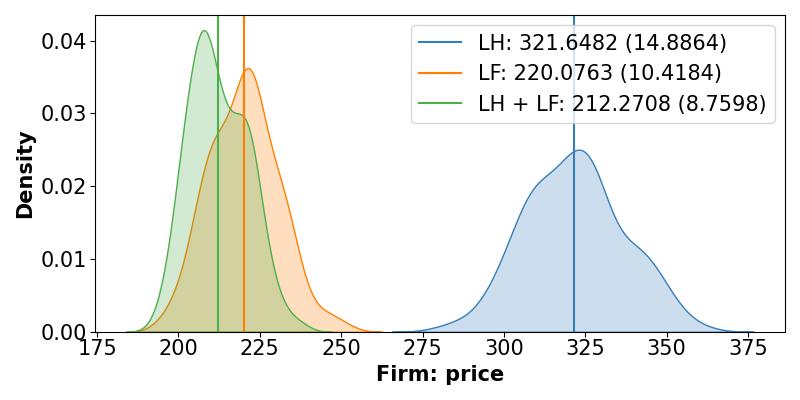}\\
    \includegraphics[width=0.45\linewidth]{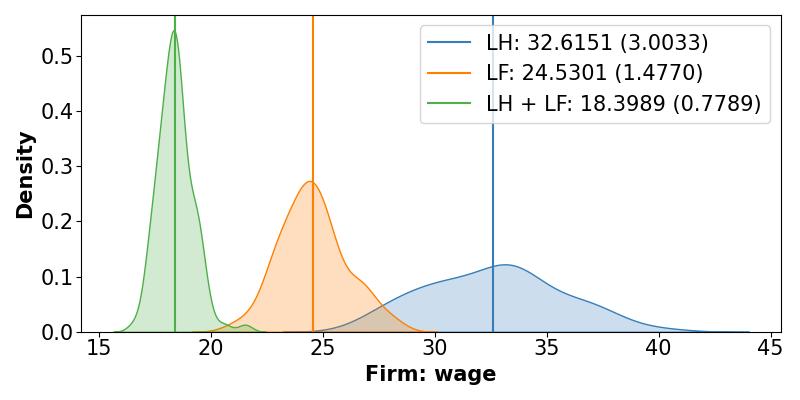}%
    \includegraphics[width=0.45\linewidth]{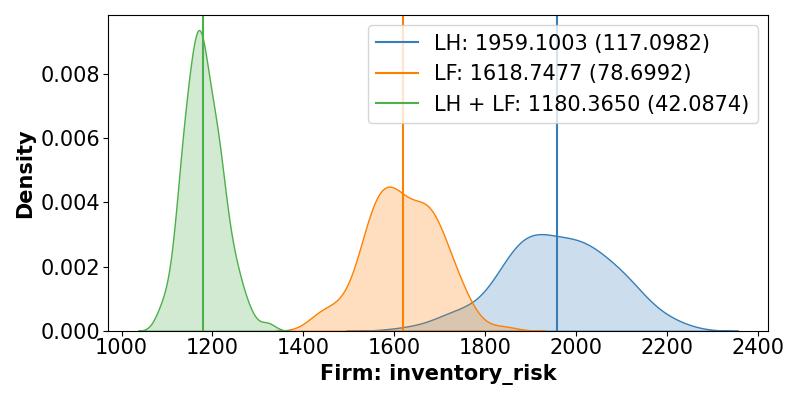}\\
    \includegraphics[width=0.45\linewidth]{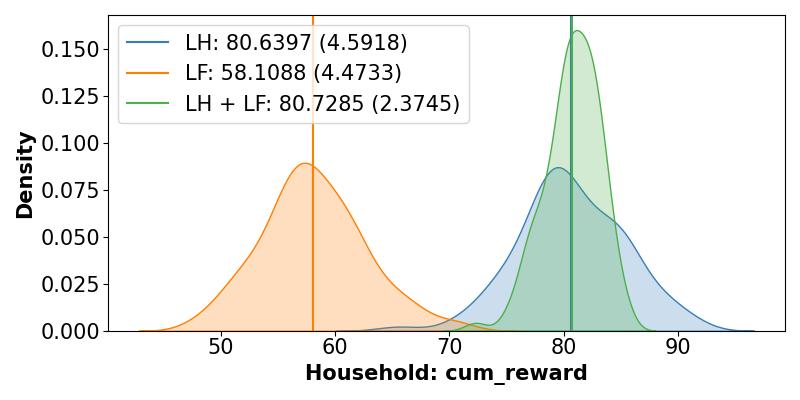}%
    \includegraphics[width=0.45\linewidth]{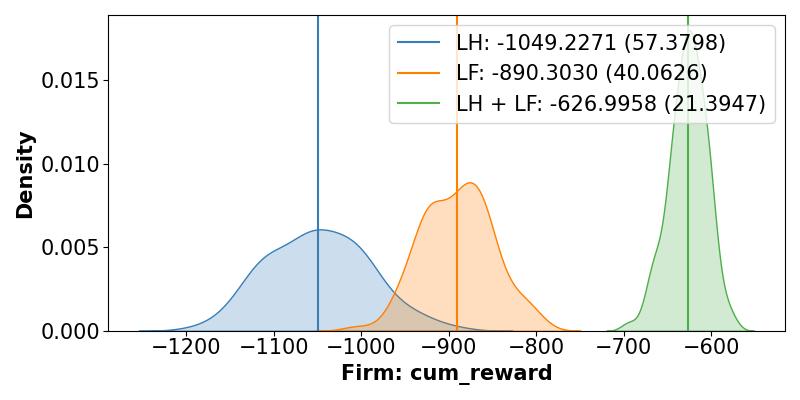}
    \caption{Histograms of key household and firm variables to demonstrate the impact of multiple learning agents. 
    Notice the increase in household consumption, decrease in firm wages, and decrease in firm prices when one or both agents are learning. Importantly, both agents achieve higher rewards when both are learning, as they adapt their behavior in response to one another.}
    \label{fig:mal_dist}
\end{figure}

To evaluate the impact on learned policies, we play out the three sets of learned policies in 100 test episodes each. Figure \ref{fig:mal_dist} shows the distribution of key household and firm observables across test episodes for the three configurations. The legends display the mean values, with standard deviations indicated in brackets. 
When the household is learning, it maximizes consumption to the highest level allowed as it can sustain savings while achieving positive consumption and savings utilities. When the firm is learning, it reduces wages to maximize profits thereby reducing household income. At the same time, it also lowers prices to maintain household consumption and minimize inventory risk.
Comparing the three configurations, we observe that when only one agent is learning, its rewards are higher than when only the other agent is learning. For example, the household achieves higher rewards in the \textbf{LH} configuration than in \textbf{LF}, while the firm achieves higher rewards in \textbf{LF} than under \textbf{LH}. Interestingly, both the household and firm achieve higher rewards when both are learning, as they optimize their own objectives while also learning to respond to each other. This is also affected by the choice of random policies for the non-learning agents in all configurations. That is, in \textbf{LF}, the household follows a random policy whereas in \textbf{LH + LF}, it learns a policy that increases consumption in presence of the firm that learns a policy to further reduce prices. This reduces the inventory risk and improves the reward of the firm alongside the household reward from consumption.

% \subsubsection{Important considerations for learning agents}
% \begin{itemize}
%     \item Exploration 
%     \item Stability and Convergence
%     \item Reward design
%     \item Generalization and Overfitting
%     \item Scalability and Computational Efficiency
% \end{itemize}
% \todo{Learning vs Non-learning agents. Below we talk about agents with capacity to learn.; how far can you push learning - more learning agents, harder to learn; scaling up etc}

\subsection{Sources of randomness}
There are two primary sources of randomness in each simulation run of our economic system. The first pertains to the employment process. At the start of each simulation episode, households are randomly assigned to firms for initial employment. Over the course of the episode, firms adjust their workforce by hiring or firing households based on their desired production levels to meet demand. When hiring, they seek unemployed households with high skill and when firing, they lay off employees with low skills. The second source of randomness arises from the exogenous shock that impacts firm production.

A subtle related point is that during training, households are assigned random skills in every episode. This ensures that policies are trained on a diverse set of observations encompassing varied skill profiles. Similarly, firms are assigned random production elasticity parameters in every episode. During testing, these skill levels and production elasticity parameters are fixed across multiple simulation episodes to evaluate policy behavior under specific scenarios. However, the randomness associated with initial employment assignments and exogenous shocks remains, introducing variability to test outcomes across episodes and providing insights into policy robustness.

% {\color{blue}SV: I think it is important to discuss need for calibration and realism before going into stylized facts} \new{-- added below.}

\section{Calibration and Stylized Facts}\label{sec:stylized_facts}

The primary objective of any model representing a real-world system is to approximate the true data-generating process accurately enough to provide a faithful representation of the data \cite{fagiolo2019validation}. For a model to be effective in forecasting system behavior or conducting policy analysis, it must closely mirror the real-world system under study.

Agent-based economic models view the system as evolving from interactions among individual actors, whose modeling is informed by empirical observations of real decision-makers. Compared to general equilibrium models, ABMs offer greater flexibility in capturing heterogeneity among economic actors, such as households and firms, and in modeling agent adaptation to changes in the economic system and environment. However, a key barrier to their widespread adoption by economists is the perceived lack of robustness, stemming from the complex relationship between ABMs and empirical data \cite{windrum2007empirical}. Despite realistic assumptions and agent descriptions, ABMs often suffer from over-parameterization due to their many degrees of freedom. 
Establishing the realism or robustness of ABMs is further complicated, compared to dynamic stochastic general equilibrium (DSGE) models, due to the non-linearities, randomness in individual behaviors and interactions, and feedback between micro and macro levels.

Existing validation approaches for ABMs fall into three categories: (1) the indirect calibration approach of comparing simulated and real world data using \textit{stylized facts} \cite{fagiolo2019validation}, (2) the Werker-Brenner approach of calibrating model parameters and initial conditions via empirical validation of resulting outputs \cite{werker2004empirical}, and (3) the history-friendly approach of calibrating model parameters and agent decision rules via reproducing historical traces \cite{windrum2007empirical}. 
In a related context, \cite{franke2012structural} introduces the method of simulated moments (MSM) for financial ABMs, where summary statistics from time series data, termed \textit{moments}, are identified. These targeted moments capture select stylized facts of interest for model validation. For example, in the context of stock price returns, these facts include the absence of return autocorrelations and volatility clustering. The goal of MSM is to adjust the ABM's parameters so that simulated moments closely match empirical moments for selected stylized facts or targeted moments. Moments not used for validation are untargeted moments, and the choice of targeted moments depends on the economic or financial scenario being simulated.

The indirect calibration approach is the most widely used given the over-parameterization issues common in ABMs, coupled with the requirement for high-quality empirical data, especially at micro-levels. 
Indirect calibration involves four steps: (1) identifying real-world stylized facts relevant to the economic scenario, (2) specifying the model, including dynamical equations for individual agents' behavior and system evolution, (3) validating the model by comparing its output with real-world data, and (4) optionally using the validated model for policy analysis. 
We examined the dynamics governing individual agents and their interactions in section \ref{sec:maes}. In this section, we provide a comprehensive list of stylized facts at various economic granularities, a subset of which may be targeted for specific problems of interest. 
We calibrate agent parameters and allowed actions using real U.S. economic data as detailed in Tables \ref{tab:parameters} and \ref{tab:action_spaces}.
We validate the model in section \ref{subsec:expt_stylized_facts} and subsequently use the validated model for policy analysis in Section \ref{subsec:expt_utility}.

Stylized facts are empirical regularities observed in real economic data, useful for validating economic models \cite{dosi2005statistical}. These patterns that persist over time can vary by scale, ranging from macroeconomic and business cycle related facts to microeconomic facts at the firm and household levels \cite{dosi2019more}. 
We survey and categorize stylized facts into macroeconomic facts that relate to aggregate macroeconomic variables such as the Gross Domestic Product (GDP), consumption, inflation, interest rate, unemployment rate, etc.; and microeconomic facts that concern distributions of household and firm variables within a region or country.

\subsection{Macroeconomic Stylized Facts}\label{subsec:macro_facts}

\subsubsection{Business Cycle Facts}\label{subsubsec:bc_facts}

We adopt the definition of a business cycle from \cite{burns1946measuring}, which describes it as:
\textit{“Consisting of expansions occurring at about the same time in many economic activities, followed by similarly general recessions, contractions, and revivals which merge into the expansion phase of the next cycle.”}
According to this definition, business cycles are recurrent but not periodic, with durations ranging from more than one year to ten or twelve years. Importantly, these cycles are indivisible into shorter cycles of a similar character. 

Stylized facts about macroeconomic time series data related to business cycles involve observed empirical relationships between Gross Domestic Product (GDP) and other economic variables at business cycle frequencies \cite{stock1999business}.
The cyclical components of macroeconomic time series refer to movements within the range of periodicities associated with business cycle durations. Extracting these cyclical components has been extensively studied, with significant contributions on appropriate methodologies and potential pitfalls \cite{hodrick1997postwar,harvey1993detrending,baxter1999measuring}.

In this work, we use the band-pass filtering technique from \cite{baxter1999measuring} to decompose macroeconomic time series into trend, cyclical, and irregular components. Then, we calculate the cross-correlation between the cyclical components of various economic series and GDP, which serves as a proxy for the business cycle. Economic series with large positive correlations with GDP are said to be pro-cyclical, while those with large negative correlations are counter-cyclical.
This methodology was employed by \cite{stock1999business} to analyze U.S. economic data from 1953 to 1996, revealing the following empirical relationships:

% Whether a series leads or lags the business cycle is determined by the maximum correlation across the different lags. For example, a maximum correlation at lag $-2$ indicates that the cyclical component of the series tends to lag the aggregate business cycle by two quarters. 
\begin{itemize}
    \item Pro-cyclical consumption expenditure
    \item Counter-cyclical prices (Consumer Price Index level)  
    \item Pro-cyclical investment in equipment
    \item Pro-cyclical inflation rate 
    \item Pro-cyclical sectoral employment 
    \item Pro-cyclical total employment  
    \item Counter-cyclical unemployment rate  
    \item Pro-cyclical total labor hours
    \item Pro-cyclical average labor productivity 
    \item Counter-cyclical nominal wages 
    \item Low correlation of real wages 
    \item Pro-cyclical nominal interest rate  
    % \item Pro-cyclical investment in equipment and nonresidential structures 
    \item Pro-cyclical imports
    \item Counter-cyclical trade balance
    \item \textit{(Phillips' Curve)} Negative correlation between cyclical components of unemployment and inflation rate
\end{itemize}
Among these, the Phillips' Curve, first proposed by \cite{phillips1958relation}, is a widely studied relationship between unemployment and inflation. While Phillips observed a long-run negative relationship between unemployment and inflation in U.K. data, \cite{stock1999business} found no stable long-run relationship between these variables in U.S. data. However, they did observe a negative correlation between their cyclical components, which supports the inclusion of this relationship among business cycle facts.

\subsubsection{Empirical regularities unrelated to Business Cycles}\label{subsubsec:macro_other_facts}
We now list empirical regularities associated with macroeconomic variables over the long run (unrelated to business cycle durations).
\begin{enumerate}
    \item \textit{(Phillips' Curve)} A negative relationship between unemployment and the rate of change of nominal wages except when there is a rapid rise in import prices (indicating a shock or war regime), as observed in economic data for the U.K. over 1861-1957 \cite{phillips1958relation}. The Phillips' curve also expects a negative relationship between unemployment and inflation rate of prices, so that low unemployment is associated with high inflation of prices and vice versa. As mentioned previously, \cite{stock1999business} do not find a stable relationship between unemployment and rate of change of wages in the U.S., but do find a negative relationship between their cyclical components.
    \item \textit{(Okun's Law)} A negative relationship between rates of change of unemployment and real GDP (expressed in percentage points), as first observed in data for the U.S. for 1947 - 1960 \cite{okun1963potential}. While this paper estimated that a 1 percentage point increase in unemployment would be associated with a 3.3 percentage point decrease in GDP, this relationship including the magnitude of which have been questioned over the years \cite{knotek2007useful,ball2017okun}.  
    \item \textit{(Beveridge Curve)} An inverse relationship between the unemployment rate and the job vacancy rate (defined as the fraction of vacant jobs relative to the labor force size), as evidenced in monthly U.S. economic data from 1952 to 1988 \cite{diamond1989beveridge,shimer2005cyclical}. This relationship is typically represented graphically, with the vacancy rate on the vertical axis and the unemployment rate on the horizontal axis. An outward shift of the curve over time indicates a scenario where the same level of vacancies is associated with higher unemployment, reflecting decreased labor market efficiency.
    % \item (Wage Curve.) A rise in unemployment is accompanied by a fall in wages when relationship is computed for regional mini-economies \cite{blanchflower1995introduction}.
    \item \textit{(Kaldor's Stylized Facts on Economic Growth)} Nicholas Kaldor proposed a set of six stylized facts to describe long-term patterns observed in economic growth, focusing on aggregate production, capital, and income distribution \cite{kaldor1961capital,jones2010new,arroyo2015stylized}. These facts were formulated as `stylized' summaries intended to capture general tendencies, not tied to precise historical accuracy but informed by economic data in the U.K. and U.S.. They are as follows:
    \begin{enumerate} 
    \item Steady growth in aggregate production and labor productivity over time, with no observable trend of diminishing productivity growth.
\item Steady growth in amount of capital per worker over time, irrespective of the specific measure of capital used.
\item A stable rate of profit on capital, consistently higher than the long-term risk-free rate.
\item Stability in the capital-GDP ratio over time, reflecting proportional growth in production and capital stock.
\item Stable shares of capital and labor in GDP, suggesting proportional growth in real wages and productivity.
\item Variations in GDP growth and labor productivity rates across countries, with faster-growing economies experiencing rates in the range of 2–5\%. 
\end{enumerate}
\end{enumerate}

\subsection{Microeconomic Stylized Facts}\label{subsec:micro_facts}
\subsubsection{Household Facts}\label{subsubsec:household_facts}
Here, we investigate the distribution of household variables such as income, wealth, and consumption across households within a region or country. 
Household income and wealth distributions exhibit significant inequality, characterized by right-skewed, fat-tailed distributions. This indicates that a small fraction of households control a disproportionately large share of income and wealth \cite{wolff1987estimates,piketty2003income}. Fat tails reflect the slower decay of the upper end of the distribution compared to exponential or normal distributions, emphasizing the concentration of resources among a minority.

Using data from the Survey of Consumer Finances, \cite{kuhn20162013} analyzed household income, earnings, and wealth distributions in the U.S. from 1989 to 2013. Metrics such as the Gini coefficient, the coefficient of variation, and Lorenz curves were employed to quantify and visualize inequality\footnote{Lorenz curves provide a visual representation of inequality by plotting the cumulative share of income (or wealth) against the cumulative share of the population, sorted by income. The degree of divergence from the 45-degree line of perfect equality underscores the concentration of resources among a small fraction of households.}.
Similarly, \cite{piketty2018distributional} constructed micro-files of pre-tax and post-tax household income in the U.S. spanning 1913 to 2014, combining tax, survey, and national account data to align microeconomic observations with macroeconomic aggregates. \cite{benhabib2018skewed} reviewed theoretical and empirical studies on household wealth distributions, exploring the economic mechanisms that drive skewness and fat tails in wealth distribution.

These studies reveal the following key stylized facts about household income and wealth distributions:
\begin{itemize}
    \item Distribution of household income is right-skewed and fat-tailed.
    \item Distribution of household wealth is right-skewed and fat-tailed. 
    \item Distribution of household wealth is more fat-tailed than income.
    \item Post-tax income is more equally distributed than pre-tax income
    \item Over time, the share of income held by the bottom 50\% of households has declined relative to the entire economy.
    \item Over time, the share of income held by the top 1\% of households has increased relative to the entire economy.
\end{itemize}
where bottom 50\% of households corresponds to those households whose incomes are lower than or equal to the median income. Similarly, the top 1\% of households corresponds to those households whose incomes lie in the 99 - 100 percentiles of the income distribution. 
% \cite{piketty2014inequality,saez2016wealth,autor2014skills} 

\subsubsection{Firm Facts}\label{subsubsec:firm_facts}
Here, we investigate the distribution of firm-related variables such as firm size, production output, profits, productivity, and growth. The statistical properties of firm size and growth rate distributions have long been the focus of empirical research \cite{gibrat1931sirey,bottazzi2003common,parham2023facts}. 
A comprehensive survey of these stylized facts is provided by \cite{dosi2005statistical}, which draws on data from U.S. manufacturing (Compustat data \cite{compustat_data}) and Italian manufacturing (ISTAT data \cite{istat_data}). The key stylized facts identified in this literature include:
\begin{enumerate}
    \item Distribution of firm sizes within sectors is right skewed, with inter-sectoral differences in firm size distributions.
    \item Distribution of firm growth rates is fat-tailed. 
    \item Decrease in variance in firm growth rates with increase in firm size.
    \item Widespread profitability differences across firms within each sector.
    \item Distribution of rate of change of firm profitabilities per sector is fat-tailed.
    \item Productivity heterogeneity across firms within each sector. 
    % \item Positive correlation between gross operating margins and labor productivity.
\end{enumerate}
where the following definitions hold:
\begin{itemize}
    \item Firm size is defined as the value added from sales (i.e., consumption spending from all households at this firm).
    % minus the cost of raw materials) / sales / number of employees
    \item Firm growth rate is defined as the change in the logarithm of firm size from step $t$ to step $t+1$.
    \item Firm profitability is defined as the difference between value added and labor costs.
    \item Firm productivity is defined as the value added per employee labor hour.
\end{itemize}

\subsection{Other Empirical Patterns.}
\paragraph{Household Finance.}
The field of household finance aims to understand the theory and empirics underlying household financial decisions \cite{campbell2006household}. A key focus is on how households make savings and consumption decisions, which influence their participation in liquid and illiquid asset markets, borrowing choices, engagement with insurance markets, and retirement savings strategies \cite{gomes2021household}.
\cite{beshears2018behavioral} provides a summary of empirical regularities in household financial behavior, particularly in the areas of consumption and savings, borrowing, asset allocation, and insurance. Drawing primarily on U.S. data, the analysis offers insights into how households manage their finances over their lifetimes. Notable regularities include:
\begin{itemize}
\item Income-consumption co-movement, with consumption expenditures closely following expected and unexpected income changes. And, expenditures declining upon retirement.
\item Low accumulation of liquid wealth, such as monetary savings, over the life cycle.
\item High incidence of credit card borrowing.
\item High accumulation of illiquid wealth, such as housing, retirement accounts, and life insurance policies, over the life cycle.
\item Increased stock market participation with increase in household wealth.
\end{itemize}

% For liquid/illiquid assets, look at percentage of households within each age range that have liquid asset holdings that are at least as large as one month of labor income. Also, look at average ratio of liquid assets to all assets across households. 
% \cite{angeletos2001hyperbolic} develop a household model with hyperbolic temporal discoutning that can account for hig credit card debt and low leves of lqiudi asets, than that can be captured by exponttial temporal discoutning.

% \paragraph{The Law of Demand.} The law of demand states that consumption of a good decreases as the price of the good increases given that other factors remain the same \cite{hildenbrand1983law}. 

\section{Experimental Results}
We begin by detailing the training setup used to learn policies for all economic agents through independent multi-agent reinforcement learning within ABIDES-Economist. 
We train agent policies under two primary economic configurations. The first configuration is aimed at validating our simulated data's ability to replicate key stylized facts outlined in Section \ref{sec:stylized_facts}. The second configuration is designed to showcase the utility of our simulation platform in crafting and comparing monetary (central bank) and fiscal (government) policies. 

For each configuration, we begin by confirming the improvement and approximate convergence of training rewards. We then evaluate the learned policies across various economic scenarios. These scenarios are crafted to either demonstrate the policies' capability to generate simulated data that aligns with targeted stylized facts among those outlined in Section \ref{sec:stylized_facts}, or to illustrate the simulator's effectiveness in developing superior monetary and fiscal policies compared to baseline rule-based approaches. 

For instance, we compare the performance of two central bank policies: the learned central bank policy, which was present during the training of other agents, and a baseline Taylor rule policy, a widely recognized benchmark in economic literature \cite{taylor2010simple}. The Taylor rule policy is introduced as an unforeseen change, allowing us to highlight the differences between the two monetary policies, while maintaining the learned policies for other agents unchanged. These experiments are run on an AWS EC2 instance (type c5.12xlarge) with 48 vCPUs and 96GB RAM. 

% Finally, we conclude this section with a discussion of our findings, providing insights into the effectiveness and implications of the learned policies within the simulated economic environment.

% Then, we explore two specific economic scenarios to highlight the utility of our platform in generating effective policies for the regulatory bodies including the central bank and the government. In the first scenario, we replace the learned monetary policy of the central bank with a Taylor rule, a widely used benchmark in economic literature \cite{taylor2010simple}, and compare the outcomes while keeping the learned policies for other agents unchanged. In the second scenario, we substitute the learned tax policy of the government with a uniform tax rate and uniform credit redistribution. This allows us to assess the impact of our learned tax policy policy in contrast to this standard benchmark, given the same learned strategies for the remaining agents.

\subsection{Learning Setup}
To scale the training process, we employed a shared policy network for all agents of the same type, reducing the number of policies to be learned to four: one for each agent type in the system. The policy network for each type receives as input the observations detailed in Section \ref{sec:maes}, augmented with the heterogeneity parameters of agents as specified in Table \ref{tab:parameters}.
Specifically, the household policy network takes skills ($\omega_{ij}$) while the firm policy network takes production elasticity ($\alpha_j$), as inputs in addition to the observations. Unless otherwise stated, agent parameters are as specified in Table \ref{tab:parameters}. 

Each policy has a continuous observation space and a discrete action space. Table \ref{tab:action_spaces} provides an overview of the values and sources for the action spaces used in our simulator.
For households and firms, action spaces comprise a uniform grid of values centered around the default values in \textbf{bold}, while adhering to any minimum value constraints. 
For the central bank, the action space spans a uniform grid of values corresponding to the range of US Federal Funds rates observed from 1950 to 2022.
For the government, the action space includes tax rates applicable to each tax bracket in 2022.

To facilitate learning, agent observations and rewards are normalized as described in Table \ref{tab:norm_reward}. Policies were trained using the Proximal Policy Optimization (PPO) algorithm implemented in the RLlib package \cite{ppo,rllib}. Learning rates were determined via a grid search over $\lbrace10^{-5}, 2 \times 10^{-5}, 5 \times 10^{-5}\rbrace$ for each agent type, selecting the first configuration that ensured consistent reward improvement across all agent types throughout training episodes. 

\begin{table*}[tb]
    \centering
    \begin{tabular}{lllll}\toprule
        Agent & Action & Values & Source \\\midrule
        Household $i$ & $c^{\textnormal{req}}_{t,ij}$& $\lbrace0,6,\textbf{12},18,24\rbrace$ & Per capita consumption\\
        & & & of 1lb of bread\\
        & & & per week \cite{statista}.\\
        \midrule
        Firm $j$ & $w_{t,j}$ & $\lbrace7.25,19.65,\textbf{32.06},44.46,56.87\rbrace$ & Minimum wage \cite{usa_gov_min_wage} \\
        & & & and average hourly earnings\\
        & & & in May 2022 \cite{bls_avg_wage}.\\
        % & Action & $w_{t,j}$ & $\lbrace7.25,\cdots,113.46\rbrace$ & \href{https://www.usa.gov/minimum-wage#}{Minimum wage} and \\
        % &&&& \href{https://www.bls.gov/oes/current/oes_nat.htm}{Estimate of Maximum median hourly wage in May 2023}\\
        & $p_{t,j}$ & $\lbrace188,255,\textbf{322},389,456\rbrace$ & Price of bread/lb in\\
        & & & May 2022 \cite{bls_price} multiplied\\
        & & & by 200 consumable goods.\\
        \midrule
        Central Bank & $r_t$ & $\lbrace0.00250,0.01625,0.03,0.04375,0.05750\rbrace$ & Federal funds rate \cite{fr_rate}\\
        \midrule
        Government & $\tau_{t,\mathrm{H}}$ & $\lbrace0.1000,0.1675,0.2350,0.3025,0.3700\rbrace$ & Lowest to highest tax \\
        & $\tau_{t,\mathrm{F}}$ & $\lbrace0.1000,0.1675,0.2350,0.3025,0.3700\rbrace$ & brackets in 2022 \cite{irs_tax}\\
        & $f_{t,i}$ & $\lbrace1,2,3,4,5\rbrace$ then, normalized by  $\sum_kf_{t,k}$\\
        \bottomrule
    \end{tabular}
    \caption{Agent action spaces in ABIDES-Economist.}
    \label{tab:action_spaces}
\end{table*}

\begin{table*}[tb]
    \centering
    \begin{tabular}{lll}\toprule
        Agent & Reward & Normalized reward \\
        \midrule
        Household $i$ & $u\left(\sum_{j}c_{t,ij},\bar{n}\sum_{j}e_{t,ij},m_{t+1,i};\gamma_i,\nu_i,\mu_i\right)$ & $u\left(\sum_j\frac{c_{t,ij}}{\bar{c}_i},\sum_{j}e_{t,ij},\frac{m_{t+1,i}}{\bar{n}\cdot\mathrm{Avg}_j\lbrace\bar{w}_j\rbrace};\gamma_i,\nu_i,\mu_i\right)$\\
        \midrule
        Firm $j$ & $p_{t,j}\sum_{i}c_{t,ij}-w_{t,j}\sum_i\bar{n}e_{t,ij}-\chi_jp_{t,j}Y_{t+1,j}$ & $\frac{p_{t,j}\sum_{i}c_{t,ij}-w_{t,j}\sum_i\bar{n}e_{t,ij}-\chi_jp_{t,j}Y_{t+1,j}}{\bar{p}_j\cdot\sum_i\bar{c}_i}$\\
        \midrule
        Central Bank & $-\left(\pi_t-\pi^\star\right)^2+\lambda\left(\sum_jy_{t,j}\right)^2$ & $-\left(\pi_t-\pi^\star\right)^2+\lambda\left(\frac{\sum_jy_{t,j}}{\sum_j\bar{y}_j}\right)^2$\\
        & & where $\bar{y}_j=\left(\bar{n}\frac{\sum_i1}{\sum_j1}\right)^{\alpha_j}$ \\
        \midrule
        Government & $\theta\sum_il_{t,i}R_{t,i,\mathrm{H}}+(1-\theta)\sum_il_{t,i}\kappa_{t,i}$ & $\theta\sum_il_{t,i}R^{\mathrm{norm}}_{t,i,\mathrm{H}}$\\
        & & $+(1-\theta)\sum_i\frac{l_{t,i}\kappa_{t,i}}{\xi\cdot\mathrm{Avg}_j\lbrace\bar{p}_j\rbrace\cdot\mathrm{Avg}_i\lbrace\bar{c}_i\rbrace}$\\
        \bottomrule
    \end{tabular}
    \caption{Normalization of agent rewards. The value of default labor hours $\bar{n}$ is given in Table \ref{tab:parameters}, while those for consumption $\Bar{c}_i$, price $\Bar{p}_j$ and wage $\bar{w}_j$ are given by the \textbf{bold faced} values in Table \ref{tab:action_spaces}.}
    \label{tab:norm_reward}
\end{table*}

\subsection{Verification of Stylized Facts}\label{subsec:expt_stylized_facts}
The selection of stylized facts to be verified must be tailored to the specific economic scenario under analysis, similar to the moments being targeted for model validation in the MSM approach. For instance, in a macroeconomic simulation investigating the effects of monetary policy on aggregate prices and production, macroeconomic facts from Section \ref{subsec:macro_facts} are more relevant than microeconomic ones from Section \ref{subsec:micro_facts}. Even within macroeconomic facts, those related to business cycles may be more important than those concerning growth theory. 
Conversely, in scenarios aimed at assessing and designing fiscal policy and tax credit distribution for households, validating microeconomic facts - particularly those related to household income and wealth distributions - is crucial for accurately capturing household-level disparities before informing policy decisions.

\begin{figure}
    \centering
    \includegraphics[width=0.9\linewidth]{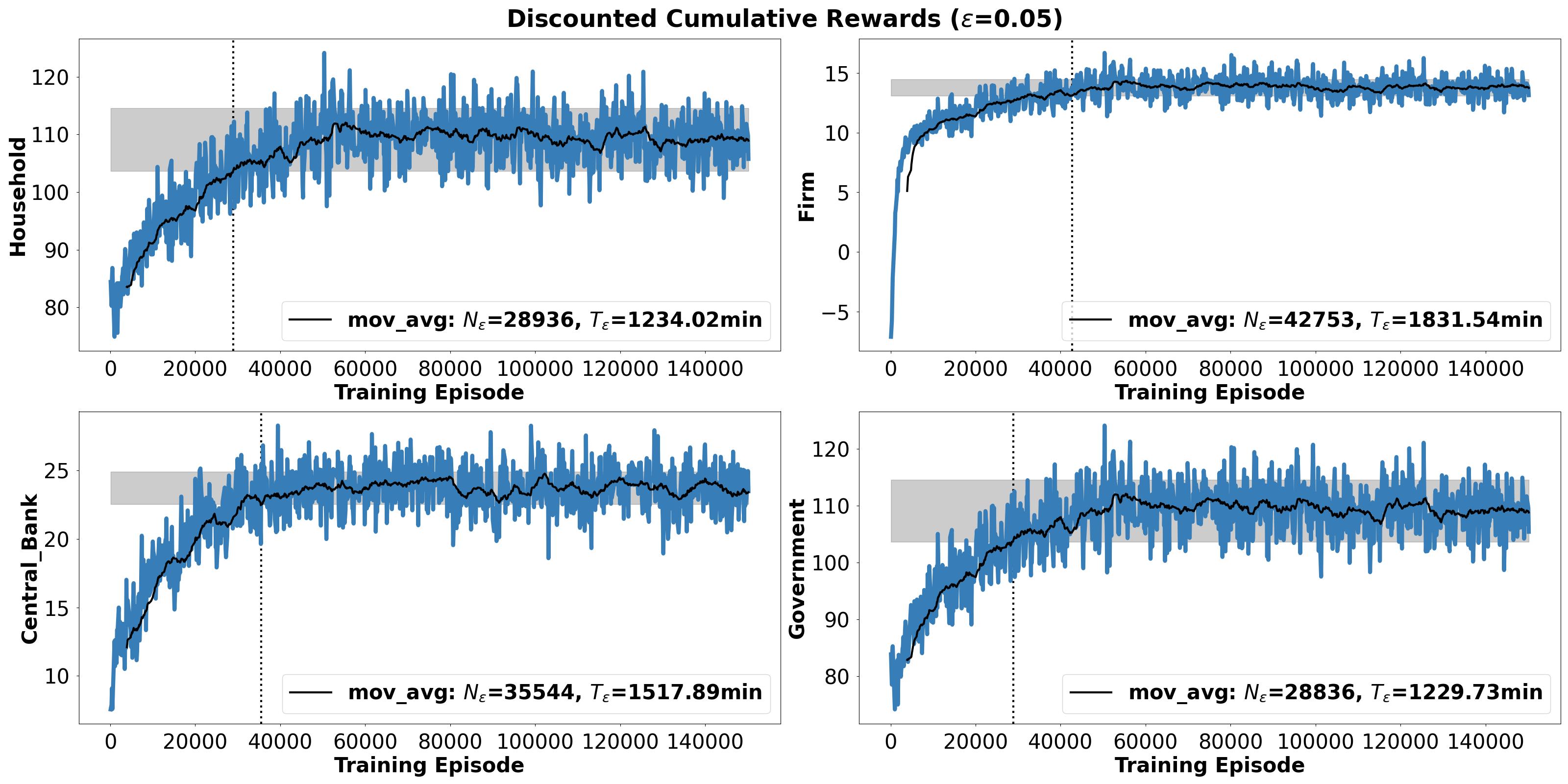}
    \caption{Discounted cumulative rewards per policy during training for the first economic configuration. Observe that moving averages of rewards converge within a $5\%$-range around long-term rewards after 42,753 episodes.}
    \label{fig:training_rewards1}
\end{figure}

For the first economic configuration, we train our economic agents across diverse regimes, including variations in household skills, firm production parameters, and exogenous shocks, as detailed in Section \ref{sec:abides_econ_sim}. Consider an economy with 100 heterogeneously skilled households, 10 heterogeneous firms, a central bank and a government as learning agents over a horizon of 10 years (40 quarters), with parameters as in Table \ref{tab:parameters}. The households share a common policy network, and as do the firms. Learning rates are set at $2\times10^{-5}$ for the household policy, $5\times10^{-5}$ for the firm policy, $5\times10^{-5}$ for the central bank policy, and $2\times10^{-5}$ for the government policy. Figure \ref{fig:training_rewards1} is a plot of discounted cumulative rewards during training for the four policies as a function of training episodes. We observe that rewards improve and stabilize beyond $5\times10^4$ training episodes demonstrating training convergence. 
In particular, it takes $N_\epsilon=42,753$ episodes (or 31 hours of training time) for the moving averages of agents' rewards to reach and remain within 5\% of the long-term rewards\footnote{Note here that we measure convergence by the moving average of rewards remaining within the $\epsilon$-percentage range around long-term rewards, rather than the rewards themselves, due to their higher variability, This increase in variability results from (1) the presence of numerous households and firms with heterogeneous parameters, and (2) the higher standard deviation of the exogenous production shock $\sigma_j$. That is to say that the training time for this configuration with 100 learning households, 10 learning firms, learning central bank and learning government is significantly higher than that observed in section \ref{subsubsec:impact_mal} when using the same instance type.}. 
% In particular, the percentage difference in the moving average of these rewards between successive episodes stays within 1\% beyond $1.4\times10^5$ episodes\footnote{See appendix for a plot of the percentage difference in the moving average of training rewards as a function of training episodes.}.

To validate the ability of our simulator to replicate stylized facts, we apply these policies across three distinct economic test scenarios. Each scenario is designed to target a specific set of stylized facts among macroeconomic facts in section \ref{subsec:macro_facts}, household microeconomic facts in section \ref{subsubsec:household_facts}, and firm microeconomic facts in section \ref{subsubsec:firm_facts}.

% we play out the learned agent policies in the following scenarios:\begin{enumerate}
%     \item Test 1: Regular economy ($\sigma_j=0.01$), with homogeneous households, typical firms with $\alpha_j\in[0.67,1]$ to demonstrate macro facts 
%     \item Test 2: Regular economy ($\sigma_j=0.01$), with heterogeneous households with skills $\omega_{ij}\sim\exp{0.66}$, typical firms with $\alpha_j\in[0.67,1]$ to demonstrate micro facts related to households
%     \item Test 3: Regular economy ($\sigma_j=0.01$), with homogeneous households, heterogeneous firms with $\alpha_j\in\textnormal{GenExtreme}{-0.05,0.28,0.15}$ to demonstrate micro facts related to firms
%     \item Test 4a: Same as in test 1 except central bank follows popular Taylor rule for interest rates to demonstrate utility of simulator to design better monetary policy
%     \item Test 4b: Same as in test 4a except with higher production process variance.
%     \item Test 5: Same as in test 1 except government follows fixed median tax rate, equal tax credit distribution to demonstrate utility of simulator to design better tax policy 
%     \item (Optional) Test 6: Same as test 4 but under positive production shock for the least labor intensive firm 
% \end{enumerate}

\subsubsection{Verification of Macroeconomic Stylized Facts}
In this macroeconomic scenario, we play out the learned policies in 100 test episodes in an economy with homogeneous households with skills $\omega_{ij}\equiv1$, homogeneous firms with production elasticity $\alpha_j\equiv0.9$\footnote{We choose $\alpha_j$ so that the production of each firm with full employment is approximately equal to the maximum consumption across households.} and low exogenous shocks $\sigma_j\equiv0.01$. 
We then extract the cyclical components of the resulting macroeconomic time series following the methodology described in Section \ref{subsubsec:bc_facts}.
\begin{figure}[h!]
    \centering
    \includegraphics[width=0.9\linewidth]{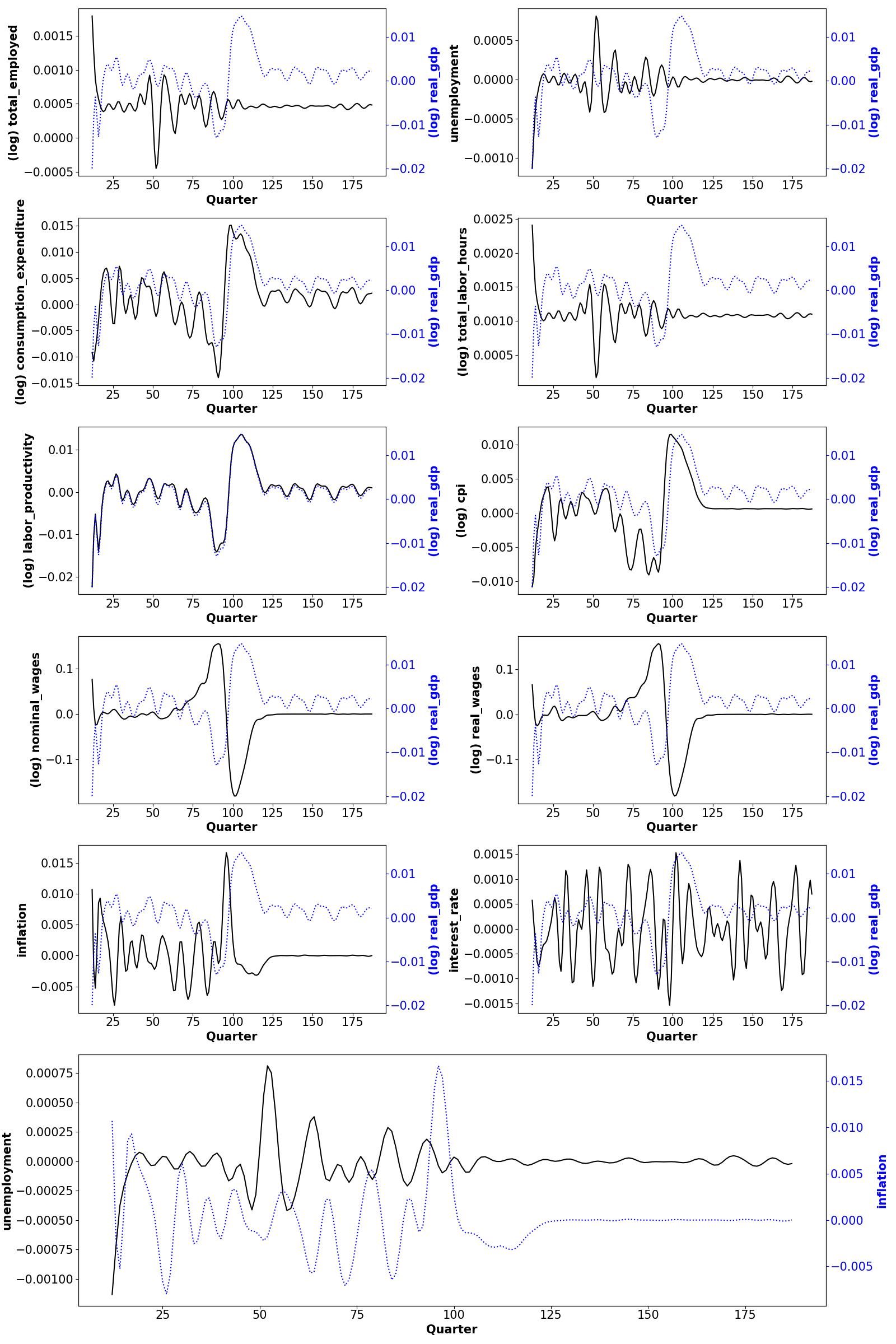}
    \caption{Cyclical components of macroeconomic variables alongside that of real GDP, unless otherwise stated. Observe that our simulated data exhibits most empirical relationships observed in real macroeconomic data related to business cycles. }
    \label{fig:macro_facts_cyc}
\end{figure}

\begin{table}[t]
    \centering
    \begin{tabular}{lr}\toprule
        Variable & Cross Correlation \\\midrule
        Total employment & 0.158642\\
        Unemployment rate & -0.158289\\
        Consumption expenditure & 0.129125\\
        Total labor hours & 0.158642\\
        Labor productivity & 0.993994\\
        Prices (Consumer Price Index) & -0.050842\\
        Nominal wage & -0.181873\\
        Real wage & -0.043230\\
        Inflation & -0.564621\\
        Interest rate & -0.014768\\
        Phillips' Curve (Unemployment, Inflation) & -0.031113\\\bottomrule
    \end{tabular}
    \caption{Cross correlations between cyclical components of macroeconomic variables with that of real GDP, unless otherwise stated. Observe that our simulated data satisfies most macroeconomic stylized facts related to business cycles. }
    \label{tab:macro_facts_corr}
\end{table}

\begin{figure}
    \centering
    \includegraphics[width=0.9\linewidth]{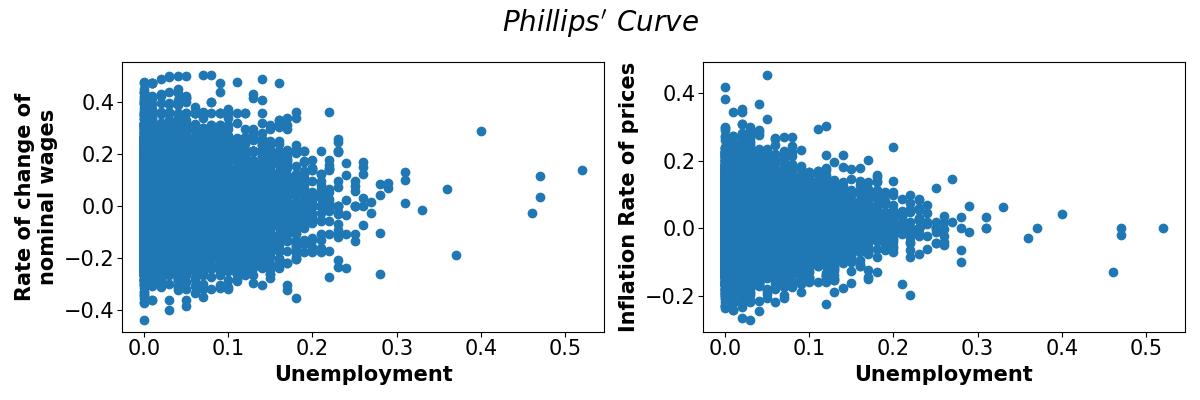}\\
    \includegraphics[width=0.45\linewidth]{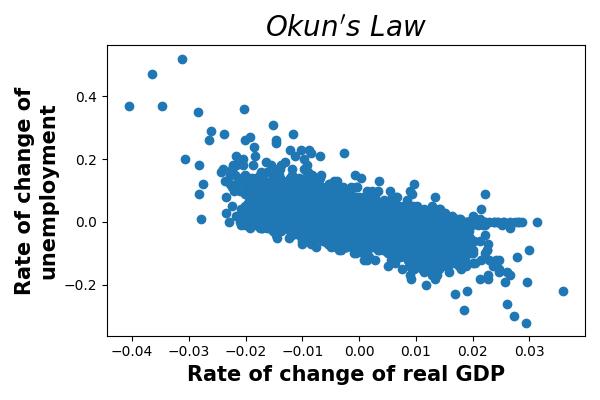}%
    \includegraphics[width=0.45\linewidth]{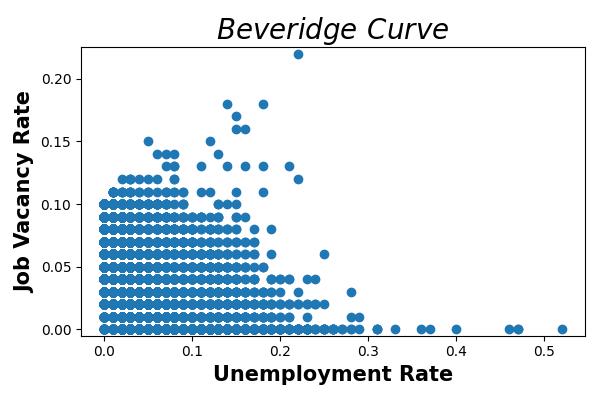}
    \caption{Validation of macroeconomic facts unrelated to business cycles.}
    \label{fig:macro_other}
\end{figure}
Figure \ref{fig:macro_facts_cyc} shows the cyclical component of key macroeconomic time series alongside that of real GDP, while Table \ref{tab:macro_facts_corr} displays their cross-correlations with real GDP. The bottom subplot of Figure \ref{fig:macro_facts_cyc} examines the cyclical components of unemployment and inflation to assess the validity of the \textit{Phillips' Curve}. Similarly, the last row of Table \ref{tab:macro_facts_corr} quantifies the cross-correlation between the cyclical components of unemployment and inflation, further validating this relationship.
We observe that the simulated data align with all the stylized business cycle facts listed in Section \ref{subsubsec:bc_facts}, except for those related to inflation and interest rates. 

We attribute the observed negative correlation between inflation and GDP to the absence of market-clearing assumptions in the goods market. Without market clearing, the mechanism driving this negative relationship is as follows: Higher GDP is associated with increased production, leading to a buildup of firm inventory. To mitigate inventory holding risks, firms lower prices, resulting in a negative correlation between inflation and GDP. On the other hand, under market-clearing conditions, higher GDP associated with higher production is matched by increased consumption. In this scenario, stronger demand drives prices upward, leading to a positive relationship between inflation and GDP.

Among the macroeconomic stylized facts unrelated to business cycles listed in Section \ref{subsubsec:macro_other_facts}, \textit{Kaldor’s Facts} do not apply to our economic model as we do not explicitly model firm capital and investment. Nonetheless, Figure \ref{fig:macro_other} illustrates relationships relevant to the \textit{Phillips' Curve}, \textit{Okun's Law}, and the \textit{Beveridge Curve}. Our findings are consistent with those observed in empirical literature. Specifically:
\begin{itemize}
    \item \textit{Phillips' Curve}: We observe no stable relationship between long-run unemployment and nominal wage changes or inflation, aligning with prior empirical findings.
    \item \textit{Okun’s Law}: We observe a negative relationship between changes in unemployment and real GDP, indicating that reductions in unemployment correspond with increases in GDP, thereby satisfying \textit{Okun’s Law}.
    \item \textit{Beveridge Curve}: We observe an inverse relationship between the job vacancy rate
    % \footnote{Here, the job vacancy rate is defined as the fraction of desired employment positions that are filled relative to the total number of households.}
    and the unemployment rate, confirming that high unemployment levels are associated with low job vacancy rates.
\end{itemize}
Thus, we successfully validate a large set of targeted macroeconomic stylized facts, including those related and unrelated to business cycles.

\subsubsection{Verification of Microeconomic Stylized Facts related to Households}
In this microeconomic scenario, we play out the learned policies in 10 test episodes in an economy with heterogeneous households with skills $\omega_{:j}\sim\mathcal{N}(0.8,0.3)$, heterogeneous firms with production elasticities linearly spaced in the range $\alpha_j\in\left[0.9,1.0\right]$, and exogenous shocks as seen during training $\sigma_j\equiv0.1$. 
The objective is to validate our simulated data on the stylized facts outlined in Section \ref{subsubsec:household_facts}, particularly the emergence of a right-skewed distribution for household income and wealth. Several models have been proposed in the literature to induce this income distribution using heterogeneities of productivity and talent \cite{benhabib2018skewed}. 
To ensure heterogeneity in income and employment outcomes, we set household skill and firm production parameters based on the following considerations:
\begin{enumerate}
    \item Since all households with skill $\omega_{ij}\geq1$ are treated equally when firms make hiring decisions, generating income heterogeneity among employed households requires wage differentiation that results from firm heterogeneity.
    \item To allow for unemployment and zero-income households in the simulation, we set higher $\alpha_j$ values and ensure that a significant fraction of households have skills below the minimum employment threshold $\omega_{\min}=1$ (see Section \ref{subsec:F}).
\end{enumerate}
% Specifically, a classical model by \todo{cite from benhabib!} assumes an exponential skill distribution among households. Here, we set the parameters of the exponential such that the probability of observing skill values outside the range of values observed during training is small\footnote{See the appendix for more information on this.}. 
\begin{figure}[tb]
    \centering
    \includegraphics[width=0.45\linewidth]{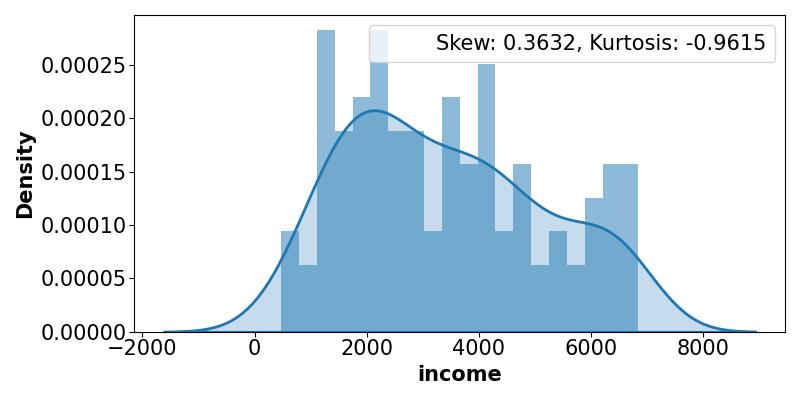}%
    \includegraphics[width=0.45\linewidth]{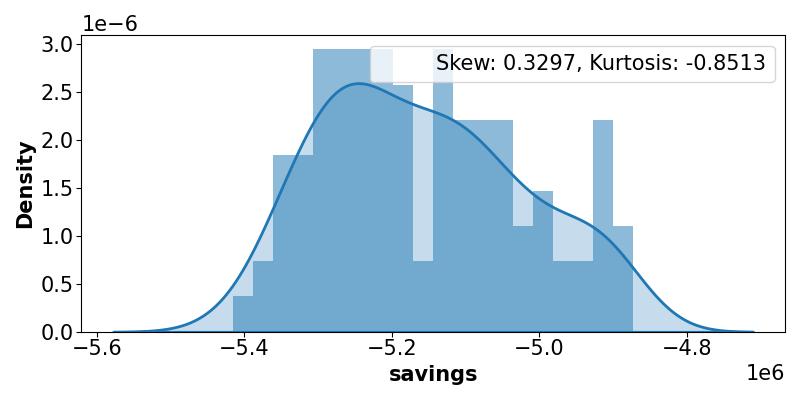}
    \caption{Distribution of average income across households (left) and final savings across households (right). Observe that income and savings are both right-skewed. }
    \label{fig:h_micro_dist}
\end{figure}

\begin{figure}[tb]
    \centering
    \includegraphics[height=1.5in]{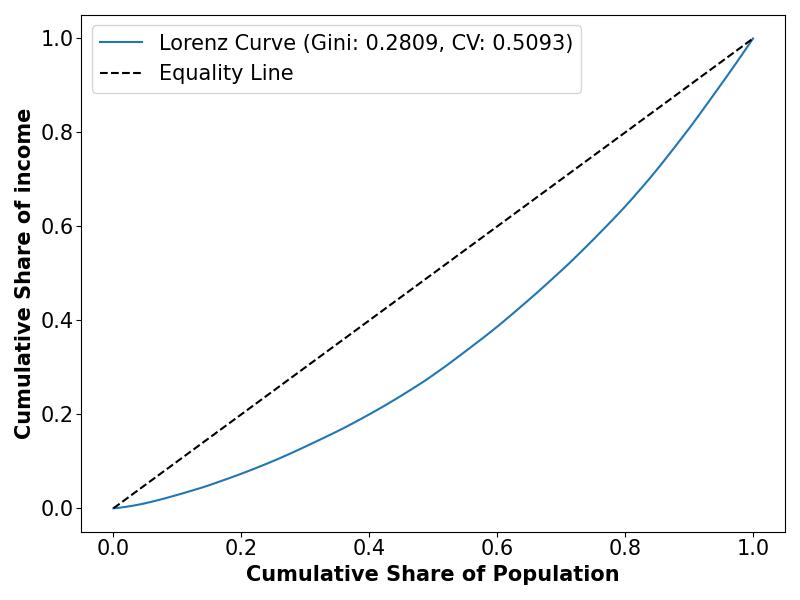}%
    \includegraphics[height=1.5in]{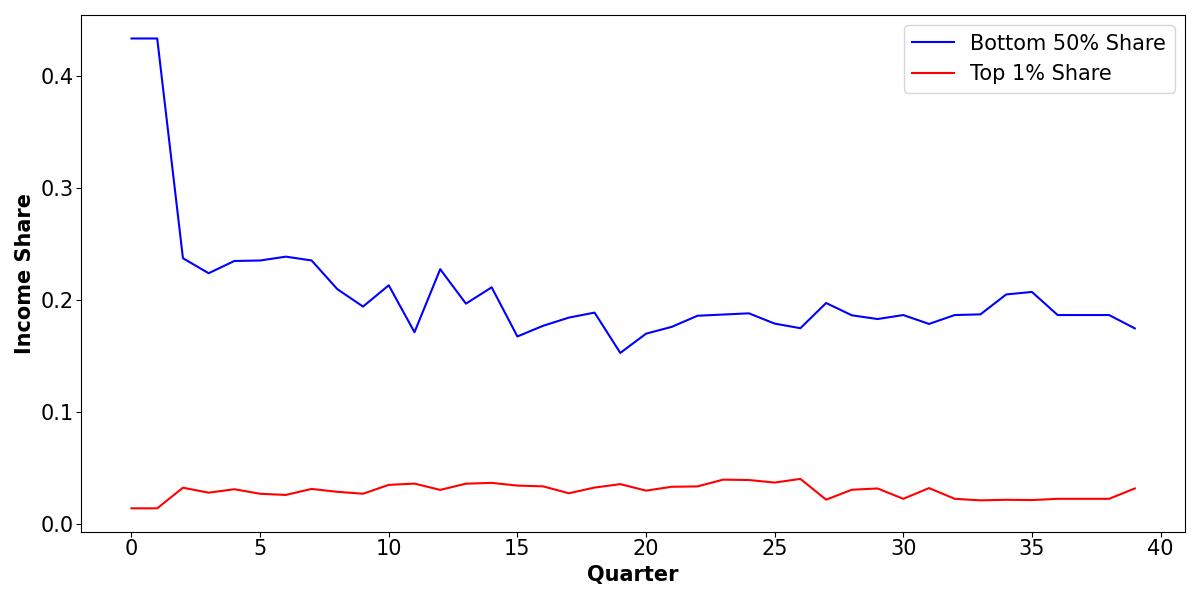}
    \caption{Visualization of household income inequality via the Lorenz curve (left) and temporal progression of income shares of the bottom 50\% and top 1\% of households (right). Observe the deviation of the Lorenz curve from the 45$^\circ$ line with positive a Gini index. Also, observe the temporal decline in the income share of the bottom 50\% of households in the economy.}
    \label{fig:h_micro_share}
\end{figure}

Figure \ref{fig:h_micro_dist} presents the distribution across households of average household income and final savings over the simulation horizon, accompanied by skewness and excess kurtosis metrics to quantify the asymmetry and fat-tailed nature of the distributions.
Both income and savings exhibit right-skewed distributions with similar levels of skewness. However, the distributions do not exhibit fat tails, as indicated by the low kurtosis values. This is due to the limited number of discrete wage levels in our simulator, which constrains the range of possible income values. Additionally, since income tax rates are constant across income brackets in our setup, the post-tax income distribution mirrors the pre-tax income distribution.

Figure \ref{fig:h_micro_share} further illustrates income inequality. The left subplot shows the Lorenz curve for household income, which visualizes income inequality by plotting the cumulative share of income against the cumulative share of the population, sorted by income. The extent of deviation from the 45$^\circ$ line of perfect equality highlights the concentration of resources among a small fraction of households. As observed in \cite{kuhn20162013}, our results show a significant deviation from perfect equality, with a large portion of the population holding a disproportionately small share of total income. This is also evident from the positive Gini index and Coefficient of variation computed on the household incomes.
The right subplot of Figure \ref{fig:h_micro_share} shows the temporal evolution of the shares of income held by the bottom 50\% and top 1\% of households. The share of income held by the bottom 50\% declines over time, while the share of the top 1\% initially increases slightly before stabilizing.

Overall, our simulated data successfully replicates the targeted microeconomic stylized facts related to households, validating the emergence of right-skewed income and wealth distributions, persistent income inequality, and the shifting of income shares between different segments of the population.

\subsubsection{Verification of Microeconomic Stylized Facts related to Firms}\label{subsubsec:expt_micro_firm_facts}
% Regular economy ($\sigma_j=0.01$), with homogeneous households, heterogeneous firms with $\alpha_j\in\textnormal{GenExtreme}{-0.05,0.28,0.15}$ to demonstrate micro facts related to firms
In this microeconomic scenario, we play out the learned policies in 10 test episodes in an economy with homogeneous households with skills $\omega_{ij}\equiv1$, heterogeneous firms with production elasticities drawn from a log-normal distribution fit to U.S. industrial data and scaled to lie within the range $\alpha_j\in\left[0.6,1.0\right]$\footnote{The scaling ensures that each firm produces at least one unit of goods per consumer household under full employment. For additional details on parameter fitting to real U.S. data, refer to the appendix.}, and low exogenous shocks $\sigma_j\equiv0.01$. 
The objective is to validate our simulated data on the stylized facts outlined in Section \ref{subsubsec:firm_facts}, those concerning the following firm-level variables:
\begin{itemize}
    \item Firm size: Measured as the total value of consumed goods $p_{t,j}\sum_ic_{t,ij}$.
    \item Growth rate: Defined as the change in log-firm size.
    \item Profitability: Measured as the difference between the total value of consumed goods and labor costs $p_{t,j}\sum_ic_{t,ij}-w_{t,j}\sum_i\bar{n}e_{t,ij}$.
    % \item Profitability rate of change: Defined as the rate of change of profitability.
    \item Productivity: Defined as the value of consumed goods per hour of employee labor $\frac{p_{t,j}\sum_ic_{t,ij}}{\sum_i\bar{n}e_{t,ij}}$.
\end{itemize}
Due to computational constraints, our experimental setup includes only 10 firms, limiting our ability to match empirical firm distributions exactly. Instead, we analyze histograms and compute skewness and excess kurtosis metrics over samples to capture distributional asymmetry and the presence of fat tails.

\begin{figure}[tb]
    \centering
    \includegraphics[width=0.45\linewidth]{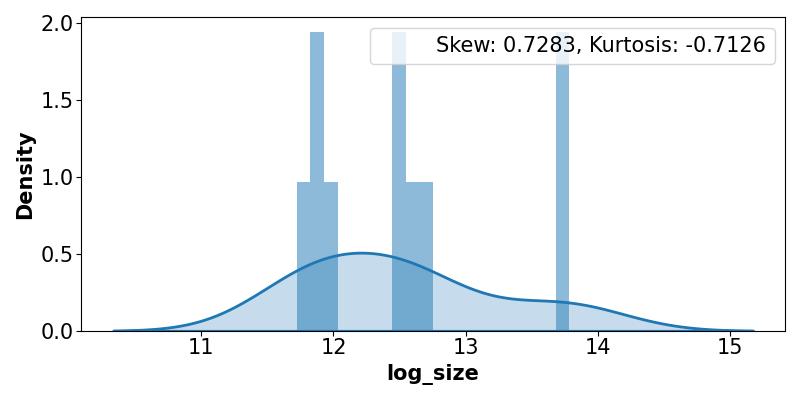}%
    \includegraphics[width=0.45\linewidth]{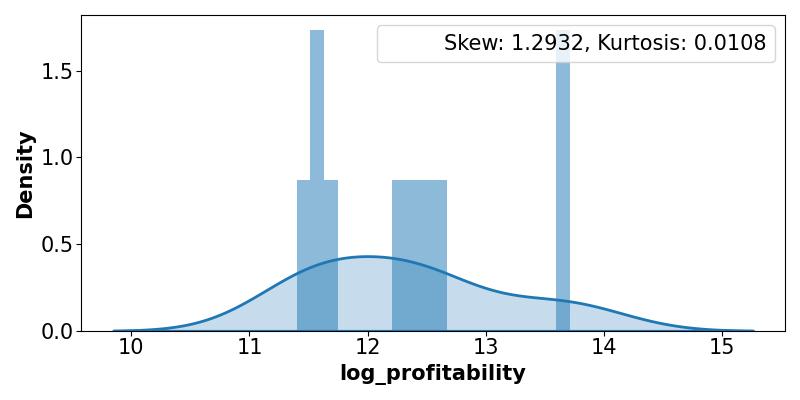}\\
    \includegraphics[width=0.45\linewidth]{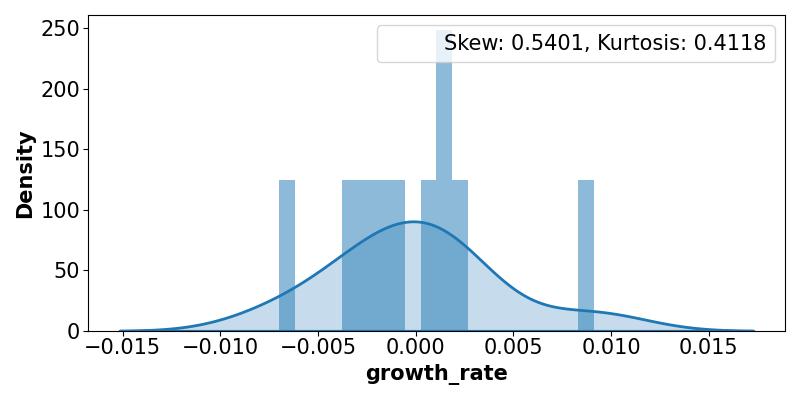}%
    \includegraphics[width=0.45\linewidth]{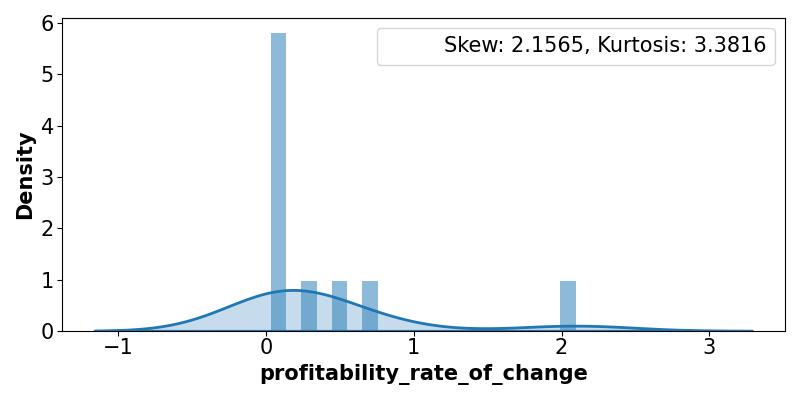}\\
    \includegraphics[width=0.45\linewidth]{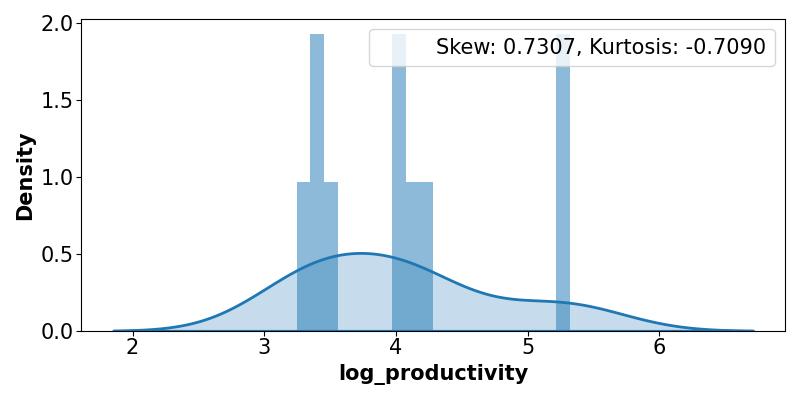}%
    \includegraphics[width=0.45\linewidth]{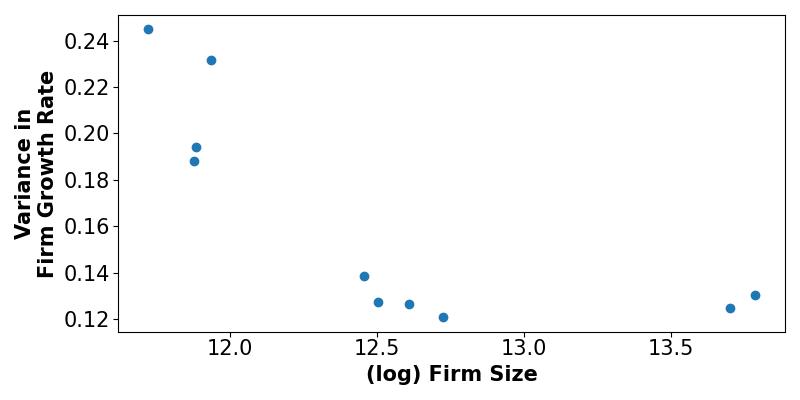}
    \caption{Histograms of key firm-related variables with skewness and excess kurtosis metrics, alongside the inverse relationship between variance in growth rate and firm size. Observe the heterogeneity across all variables, right-skewed firm size distribution, and fat-tailed distributions for growth rate and profitability rates of change.}
    \label{fig:f_micro_dist}
\end{figure}

Figure \ref{fig:f_micro_dist} presents histograms of the aforementioned variables, averaged per firm over the simulation horizon, along with the average rate of change of profitability. Figure \ref{fig:f_micro_dist} also displays the relationship between variance in firm growth rates across time steps and firm size.
Here, we observe heterogeneity across firms in all measured variables. The histogram of firm sizes exhibits positive skewness, consistent with empirical findings. The firm growth rate histogram shows positive excess kurtosis, indicating fatter tails than a normal distribution.
Likewise, the rate of change of profitability is also fat-tailed. 
Importantly, despite the small number of firms, we observe an inverse relationship between firm size and the variance of growth rates, as shown in the bottom-right subplot of Figure \ref{fig:f_micro_dist}.

Overall, our simulated data successfully replicates the targeted microeconomic stylized facts related to firm heterogeneity, though the small sample size of firms presents some limitations in matching real-world distributions exactly.

\subsection{Utility for Monetary and Fiscal Policy Design}\label{subsec:expt_utility}
We now demonstrate the efficacy of our simulation platform in aiding policy design and analysis for regulatory bodies, such as the central bank and government. 
Specifically, we aim to illustrate the impact of different regulatory policies on achieving policy objectives in the presence of heterogeneous and adaptive households and firms. 
For comparison, we utilize two central bank policies: the learned policy which was seen during the training of other agents, and the widely adopted rule-based Taylor Rule policy \cite{taylor1993discretion}, which serves as an unforeseen change. Likewise, we compare the learned government policy with a uniform tax rule on fiscal policy efficacy. These baselines are introduced as unforeseen changes to regulatory policy, as they were not encountered by other agents during their learning process.

This second economic configuration mirrors the first in capturing diverse regimes via variations in household skills, firm production parameters, and exogenous shocks, as described in Section \ref{subsec:expt_stylized_facts}. However, it introduces a key modification to the government reward parameter, specifically setting $\theta=0.2$ to place greater emphasis on tax credit redistribution over resulting household utility, as outlined in equation (\ref{eq:G_reward}). 
The choice of a different $\theta$ value, where $\theta<1$, is intentional as it aims to enhance the redistribution of tax credits towards reducing savings inequality among households. 
This adjustment influences our simulated data's ability to generate microeconomic patterns related to household inequality, as the government actively seeks to minimize discrepancies, prompting households to adapt accordingly. To address these dynamics, we explore two distinct economic configurations: (1) for the validation of stylized facts, as discussed in the previous subsection, and (2) for demonstrating the simulator's utility in policy design, as covered in this subsection.

\begin{figure}
    \centering
    \includegraphics[width=0.9\linewidth]{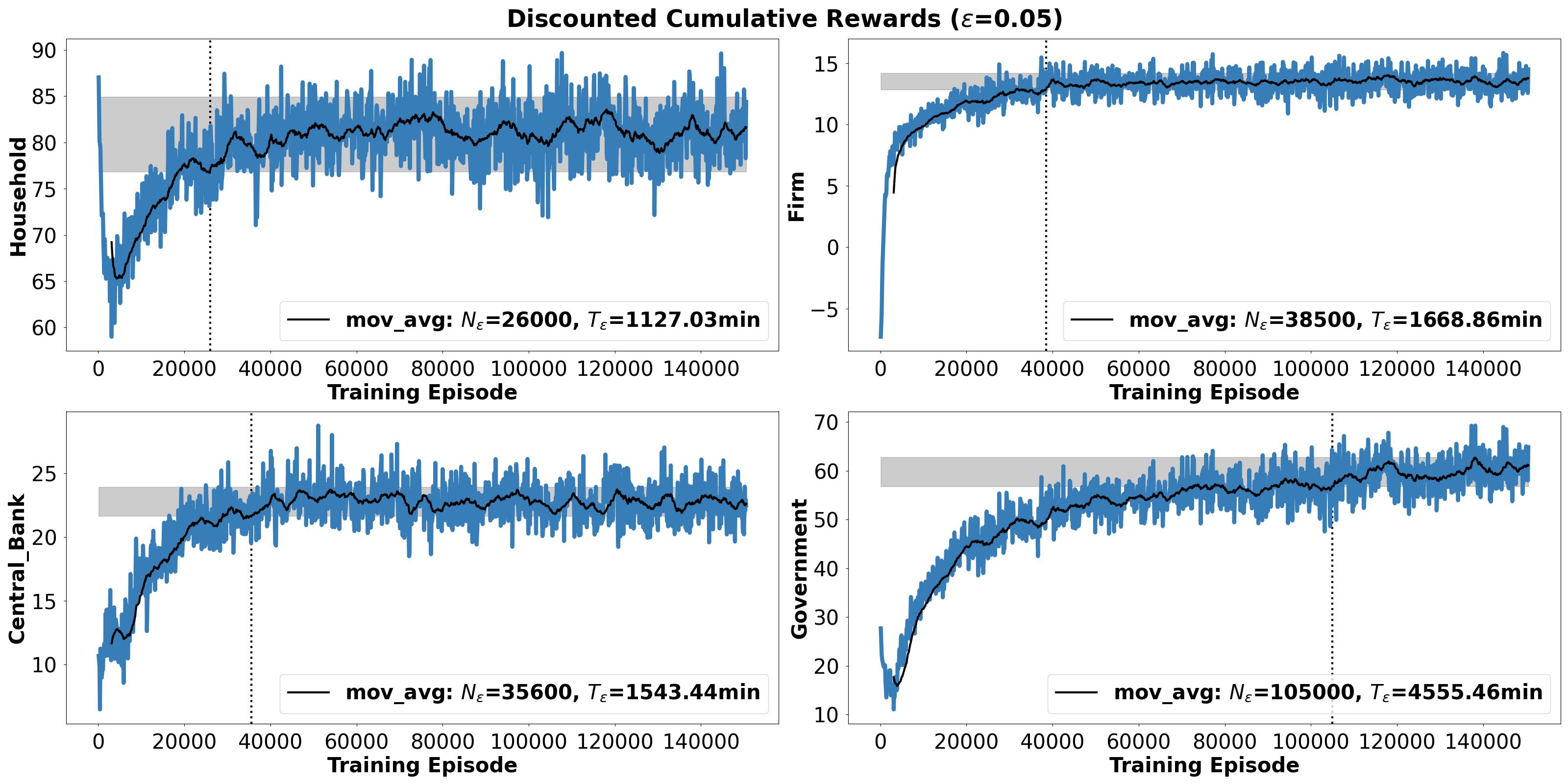}
    \caption{Discounted cumulative rewards per policy during training for the second economic configuration. Observe that moving averages of rewards converge within a $5\%$-range around long-term rewards after 105,000 episodes.}
    \label{fig:training_rewards2}
\end{figure}

For the second economic configuration, consider an economy with 100 heterogeneously skilled households, 10 heterogeneous firms, a central bank and a government as learning agents over a horizon of 10 years (40 quarters), with parameters as in Table \ref{tab:parameters} except with $\theta=0.2$. The households share a common policy network, and as do the firms. Learning rates are set at $2\times10^{-5}$ for the household policy, $5\times10^{-5}$ for the firm policy, $5\times10^{-5}$ for the central bank policy, and $2\times10^{-5}$ for the government policy. Figure \ref{fig:training_rewards2} is a plot of discounted cumulative rewards during training for the four policies as a function of training episodes. We observe that rewards improve and stabilize beyond $10^5$ training episodes demonstrating training convergence. 
In particular, it takes $N_\epsilon=105,000$ episodes (or 76 hours of training time) for the moving averages of agents' rewards to reach and remain within 5\% of the long-term rewards. 
% In particular, the percentage difference in the moving average of these rewards between successive episodes stays within 1\% beyond $1.2\times10^5$ episodes\footnote{See appendix for a plot of the percentage difference in the moving average of training rewards as a function of training episodes.}.

\subsubsection{Utility for Monetary Policy Design}
In this macroeconomic scenario, we play out the learned policies in 100 test episodes, with a focus on central bank decision-making. The economy consists of homogeneous households with skills $\omega_{ij}\equiv1$, heterogeneous firms with production elasticities linearly spaced in the range $\alpha_j\in\left[0.05,1.00\right]$, and low exogenous shocks $\sigma_j\equiv0.01$. 
Subsequently, to establish a baseline for comparison and highlight the benefits of learning a central bank policy, we substitute the learned central bank policy with a non-inertial Taylor rule for setting interest rates, as described in \cite{taylor1993discretion}:
\begin{align}
    r_{t+1}=r^\star+\pi_t+0.5(\pi_t-\pi^\star)+0.5(y_t-y_t^\star)\nonumber
\end{align}
where $r^\star=2\%$ represents the equilibrium interest rate, $\pi_t$ is the current inflation rate, $\pi^\star=2\%$ is the target inflation rate, and $y_t-y_t^\star$ denotes the output gap between current production output $y_t=\sum_jp_{t,j}y_{t,j}$ and an estimate of potential output $y_t^\star$ computed using a linear fit to past outputs.
We evaluate the performance of the learned central bank policy against the Taylor rule under identical conditions, where all other economic agents continue to use their learned policies. The discounted sum of the central bank's rewards as defined in equation (\ref{eq:CB_reward}) capture the central bank's utility from minimizing deviations from target inflation, and maximizing GDP. This serves as the performance metric for comparing monetary policies. 
\begin{figure}[tb]
    \centering
    \includegraphics[width=0.9\linewidth]{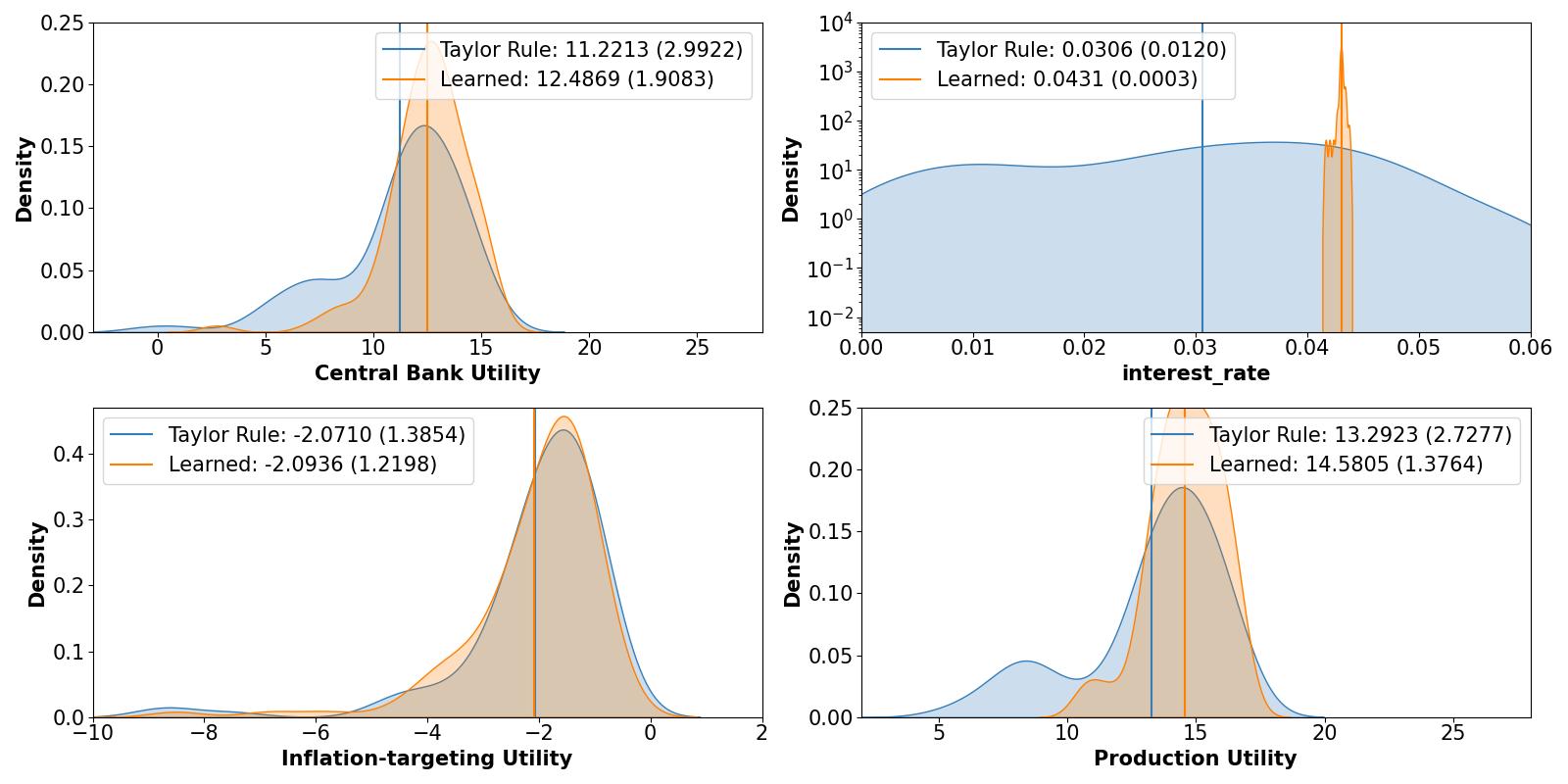}
    \caption{Distribution of Central Bank utility from inflation targeting and GDP, along with other monetary observables for the Taylor rule (\textcolor{myblue}{blue}) and learned policy (\textcolor{myorange}{orange}) in a regular economy with low production shocks. The learned policy achieves higher utility with lower variance. }
    \label{fig:utility_monetary}
\end{figure}

Figure \ref{fig:utility_monetary} presents the distribution of central bank utility and other monetary observables across test episodes. 
The results for the Taylor rule are shown in \textcolor{myblue}{blue}, while those for the learned policy are shown in \textcolor{myorange}{orange}. The legends display the average value and standard deviation (in brackets) across test episodes. We observe that the central bank achieves slightly higher utility with lower variance under the learned policy compared to the Taylor rule. In addition, while the Taylor rule utilizes nearly all interest rate options, the learned policy typically employs at most two rate options on average\footnote{This comparison is based on the average interest rate over the horizon across test episodes, so variability over the horizon is still present, as expected.}.
Furthermore, although both policies achieve similar inflation targeting, the learned policy is more effective in promoting production. 

To further assess the robustness of the learned policy, we compare its performance with that of the Taylor rule in a volatile economic scenario with high exogenous shocks to the firms. In particular, the mean of the firm shock process is linearly spaced in the range $\bar{\varepsilon}_j\in\left[0.10,0.00\right]$, with the standard deviation linearly spaced in the range $\sigma_j\in\left[0.20,0.10\right]$. 
This scenario represents an economic environment where the firm with the lowest production elasticity $\alpha_1=0.05$ experience positive production shocks with high mean and variance $(\bar{\varepsilon}_1,\sigma_1)=(0.10,0.20)$. And, the firm with the highest production elasticity $\alpha_1=1.0$ experiences low production shocks as before, with low mean and variance $(\bar{\varepsilon}_1,\sigma_1)=(0.00,0.10)$. 

\begin{figure}[tb]
    \centering
    \includegraphics[width=0.9\linewidth]{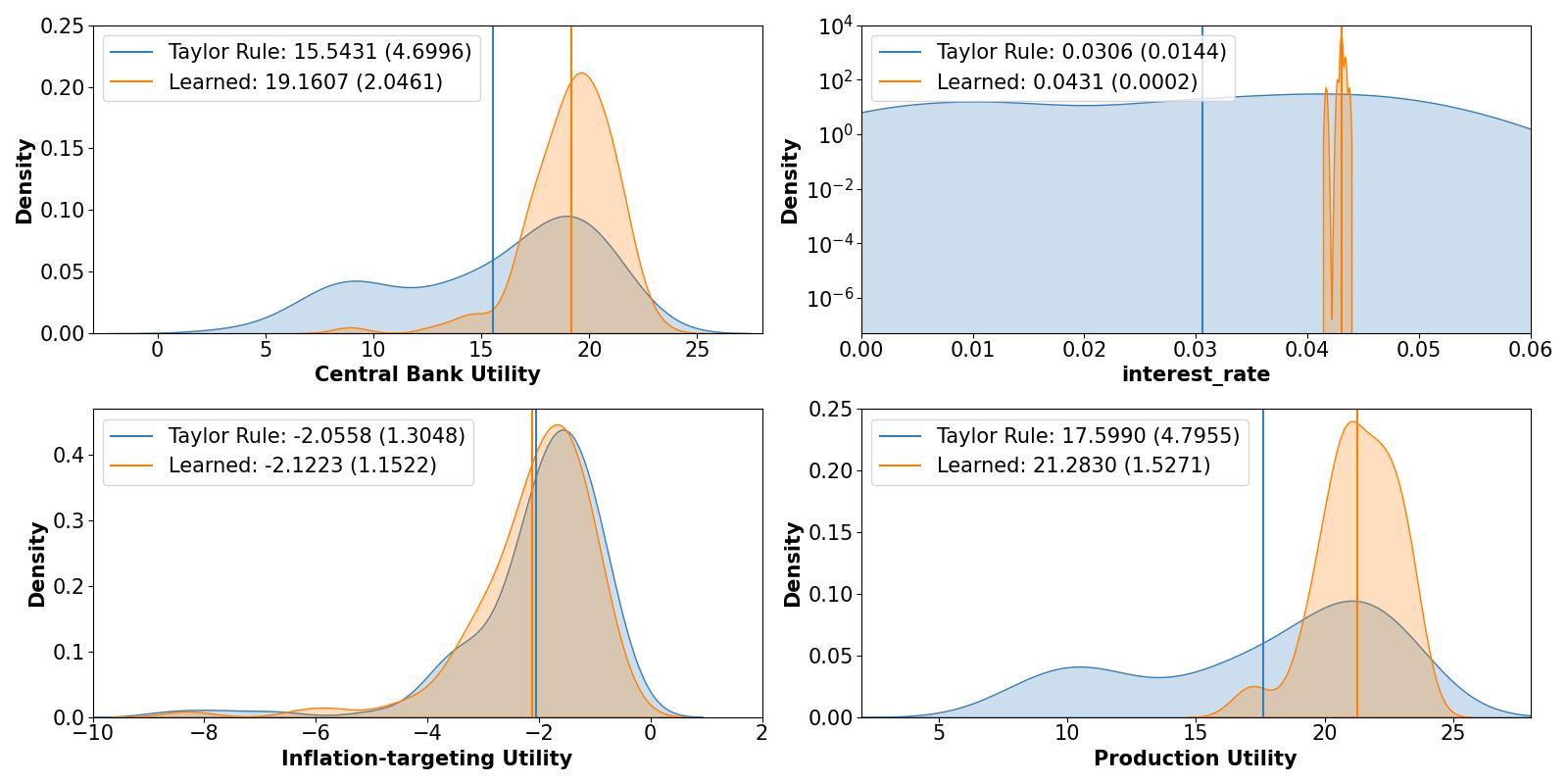}
    \caption{Distribution of Central Bank utility from inflation targeting and GDP, along with other monetary observables for the Taylor rule (\textcolor{myblue}{blue}) and learned policy (\textcolor{myorange}{orange}) in a volatile economy with high production shocks. The learned policy significantly outperforms the Taylor rule, and has lower variance that is similar to the scenario without shocks.}
    \label{fig:utility_monetary_shock}
\end{figure}

Figure \ref{fig:utility_monetary_shock} presents the distribution of central bank utility and other monetary observables across test episodes in this economic scenario characterized by high variability due to heterogeneity in firm productions and exogenous shocks. We observe that the learned policy significantly outperforms the Taylor rule in achieving higher utility for the central bank in inflation targeting and GDP promotion, while also maintaining lower variance across episodes. Notably, the variance in central bank utility for the learned policy remains similar to the previous low-shock scenario, whereas the variance for the Taylor rule is substantially higher than before. 
As previously noted, while both policies achieve similar inflation targeting objectives, the learned policy is more effective in promoting production under such shock scenarios compared to the Taylor rule. This highlights the effectiveness of the learned central bank policy in meeting inflation and production targets through strategic interest rate setting even in volatile economic scenarios.

\subsubsection{Utility for Fiscal Policy Design}
We consider a macroeconomic scenario akin to the one above for monetary policy testing, but now with a focus on government decision-making, particularly concerning heterogeneous households. We play out learned policies in 100 test episodes mirroring regimes seen during training. The economy consists of heterogeneous households with skills $\omega_{ij}\sim\mathcal{N}(1.0,0.3)$, heterogeneous firms with production elasticities linearly spaced in the range $\alpha_j\in\left[0.05,1.00\right]$, and low exogenous shocks $\sigma_j\equiv0.01$. 
Subsequently, to establish a baseline for comparison and highlight the benefits of learning a government policy for tax redistribution, we substitute the learned government policy with a fixed median tax rate and a uniform tax redistribution policy:
\begin{align}
    \tau_{t+1,\mathrm{H}}&=\tau_{t+1,\mathrm{F}}=0.2350\nonumber\\
    f_{t+1,i}&=\frac{1}{n}\ \forall i
\end{align}
where $0.2350$ is the tax rate corresponding to the mid tax bracket in Table \ref{tab:action_spaces}, and $n$ denotes the number of households. 
We evaluate the performance of the learned government policy against this uniform tax rate rule under identical conditions, where all other economic agents continue to use their learned policies. The discounted sum of the government's reward as defined in equation (\ref{eq:G_reward}), reflects its utility from household social welfare as a weighted sum of household utilities and tax credits. This serves as the performance metric for comparing fiscal policies. Note that we apply inverse-income weights for households, as detailed in Table \ref{tab:parameters}.
\begin{figure}[tb]
    \centering
    \includegraphics[width=0.9\linewidth]{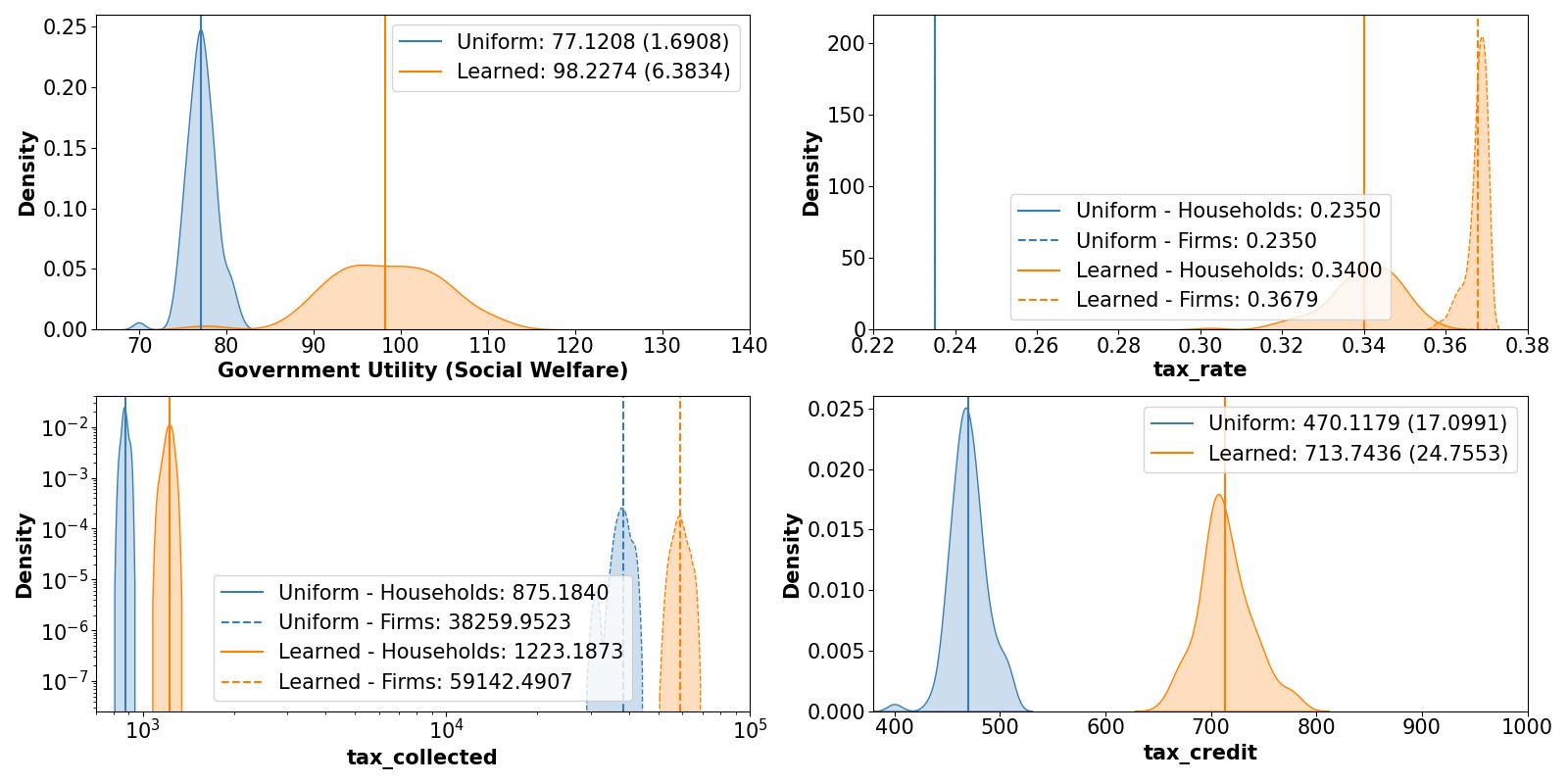}
    \caption{Distribution of household social welfare and tax related observables for the uniform tax policy (\textcolor{myblue}{blue}) and learned policy (\textcolor{myorange}{orange}) in a regular economy with low production shocks. The learned policy achieves higher social welfare via greater tax collection and redistribution than a rule-based uniform tax policy. }
    \label{fig:utility_fiscal}
\end{figure}

Figure \ref{fig:utility_fiscal} displays the distribution of social welfare, alongside the tax rates for households and firms, taxes collected from them, and the tax credits provided to households across test episodes. Results for the uniform tax rule are shown in \textcolor{myblue}{blue}, while those for the learned policy are shown in \textcolor{myorange}{orange}. The learned tax policy achieves higher social welfare compared to the uniform tax policy. It sets higher tax rates on average and collects larger amounts of taxes from both households and firms, which facilitates the redistribution of larger tax credits to households. This highlights the effectiveness of the learned government policy in enhancing social welfare through strategic tax credit distribution.

Similar to our approach with monetary policy, we further assess the robustness of the learned tax policy in a volatile economic scenario characterized by high exogenous shocks to firms. Specifically, the mean of the firm shock process is linearly spaced in the range $\bar{\varepsilon}_j\in\left[0.10,0.00\right]$, with the standard deviation linearly spaced in the range $\sigma_j\in\left[0.20,0.10\right]$. This scenario features high economic variability due to heterogeneity in firm productions and exogenous shocks, alongside the heterogeneity in household skills.

\begin{figure}[tb]
    \centering
    \includegraphics[width=0.9\linewidth]{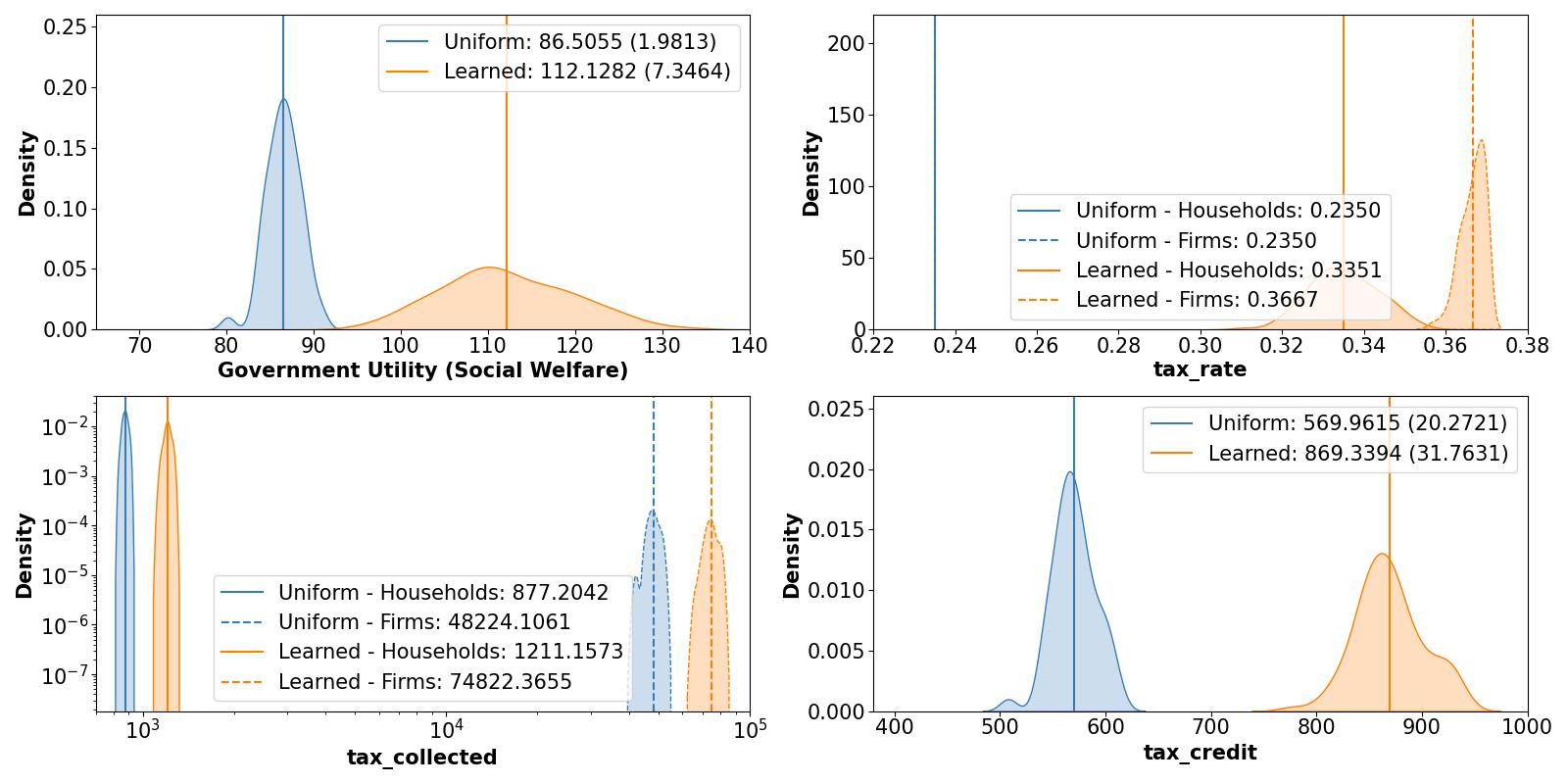}
    \caption{Distribution of household social welfare and tax related observables for the uniform tax policy (\textcolor{myblue}{blue}) and learned policy (\textcolor{myorange}{orange}) in a volatile economy with high production shocks. The learned policy achieves higher social welfare via greater tax collection and redistribution. While both fiscal policies exhibit an increase in social welfare variability due to shock volatility, the variance increase is smaller for the learned policy.}
    \label{fig:utility_fiscal_shock}
\end{figure}

Figure \ref{fig:utility_fiscal_shock} presents the corresponding distribution of social welfare and tax-related observables across the test episodes. We observe that both fiscal policies achieve slightly higher social welfare due to increased tax collection from firms, even with unchanged tax rates. This increase is attributed to higher firm profits resulting from greater production due to the positive shocks. Notably, the percentage increase in social welfare is larger for the learned policy compared to the previous low-shock scenario. The increased variability of shocks is reflected in the higher variance of social welfare for both fiscal policies compared to the low-shock scenario. Nonetheless, the percentage increase in the variance of social welfare is lower for the learned policy than for the rule-based policy. Importantly, the learned policy continues to outperform the rule-based uniform tax policy even in this volatile high-shock economy.

% \subsection{Impact of having multiple learning agents}

\section{Conclusion}
We introduce ABIDES-Economist, a multi-agent simulation platform designed for economic systems featuring heterogeneous households, firms, a central bank, and a government. Our simulator is configurable for both microeconomic and macroeconomic granularities, offering versatility for simulating diverse economic scenarios. It enables economic agents to utilize reinforcement learning algorithms to develop strategies that maximize their objectives, even in the presence of shocks, agent heterogeneity, and adaptation.

A key contribution of our work is toward calibration and validation of economic agent-based simulators with learning agents. We provide a comprehensive survey of macroeconomic and microeconomic stylized facts for households and firms. By grounding our agent parameters in real U.S. economic data where available, we validate our platform's ability to replicate a wide range of targeted micro- and macroeconomic facts, even when all agents use reinforcement learning to determine their behavioral policies. Upon validation, we demonstrate the platform's utility for economic policy-making, particularly in scenarios where agent heterogeneity and economic shocks are significant. We compare policies developed within our platform to standard rule-based approaches from the literature, illustrating the superior performance of our policies. This paves the way for studying and designing economic policies within a validated and controlled framework before real-world implementation.

While we demonstrate the platform's utility for economic policy analysis and design, we also recognize the challenges of scalability and learning instability inherent to multi-agent reinforcement learning. Our experiments showcase the capability to simultaneously learn policies for 100 households, 10 firms, 1 central bank, and 1 government. We achieve this scaling by sharing a single policy network across all agents of the same type to accelerate learning through shared experiences. Additionally, we implement realistic agent communication, where agents exchange shareable information, reducing partial observability challenges associated with multi-agent reinforcement learning.

Lastly, the quality of the equilibrium to which the learned policies converge remains uncertain. We propose incorporating game-theoretic tools suited for agent-based models, such as empirical game-theoretic analysis, to compare and identify different equilibria for multi-agent systems with learning agents \cite{wellman2020economic,lanctot2017unified}. Although this approach incurs additional computational costs due to simulating utilities for an exponentially increasing set of joint policies as the agent/policy count grows, it can mitigate instabilities in multi-agent reinforcement learning by allowing agents to iterate over unilaterally improving their policies.
\clearpage
\newpage
\backmatter

% \bmhead{Supplementary information}

\bmhead{Acknowledgements}
The authors would like to thank Dr. Jesse Perla for his insightful comments on existing literature from economics and macrofinance.
This paper was prepared for informational purposes in part by the Artificial Intelligence Research group of JPMorgan Chase \& Co. and its affiliates (“JP Morgan’’) and is not a product of the Research Department of JP Morgan. JP Morgan makes no representation and warranty whatsoever and disclaims all liability, for the completeness, accuracy, or reliability of the information contained herein. This document is not intended as investment research or investment advice, or a recommendation, offer or solicitation for the purchase or sale of any security, financial instrument, financial product, or service, or to be used in any way for evaluating the merits of participating in any transaction, and shall not constitute a solicitation under any jurisdiction or to any person, if such solicitation under such jurisdiction or to such person would be unlawful. 

\begin{appendices}

\section{Estimating Firm Productivity from Labor using Real Data}\label{sec:app:firm_alpha}
We obtained annual employment and output data for all industry sectors within the United States from the Bureau of Labor Statistics \cite{bls_ind_output}. This data includes aggregated variables related to employment and productivity for firms across 167 sectors, which are grouped into 22 industry groups. For example, the group \textit{`Agriculture, Forestry, Fishing, and Hunting'} encompasses data from six sectors, including crop production; animal production and aquaculture; forestry; logging; fishing, hunting and trapping; and support activities for agriculture and forestry.
We filtered 20 out of the 22 groups, excluding the special industries and value-added industries groups due to incomplete data. As a result, we have annual data for 163 industry sectors within these 20 groups. Specifically, we utilize annual domestic industry output in millions of current dollars from 1997 to 2023 and annual total labor hours of all employed persons in millions from 2014 to 2023. 
% This provides us with labor hours and industry output data for each sector across 20 groups over a ten-year period.
We then aggregate labor hours and employment by group, calculating total labor hours and total employment for each of the 20 industry groups over ten years. 

\begin{figure}[!h]
    \centering
    \includegraphics[width=0.45\linewidth]{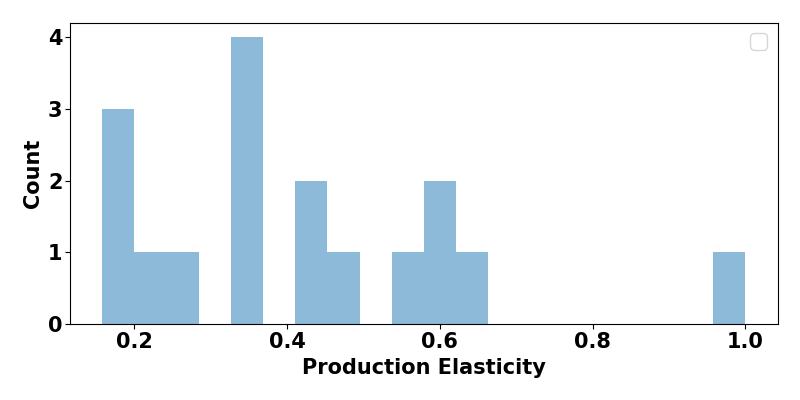}
    \caption{Histogram of production elasticities for labor, fit to employment and output data for 18 industry groups in the U.S.}
    \label{fig:f_alpha_fit1}
\end{figure}

To estimate the production elasticity for labor $\alpha_j$ for each industry group $j$, we fit a linear regression model to the log-transformed labor hours and output data:
\begin{align}(\hat{\alpha}_j,\hat{\beta}_j)\in\arg\min_{\left(\alpha,\beta\right)}\frac{1}{T}\sum_{t=1}^T\left(\log{y_{t,j}}-\alpha\log{N_{t,j}}-\beta\right)^2
\end{align}
We scale the resulting $\alpha_j$ values by the largest elasticity and exclude the two groups with negative elasticities, as we require all production elasticities to lie within $[0,1]$. This yields the production elasticities fitted to real U.S. data, as shown in Figure \ref{fig:f_alpha_fit1}.

\begin{figure}[!h]
    \centering
    \includegraphics[width=0.45\linewidth]{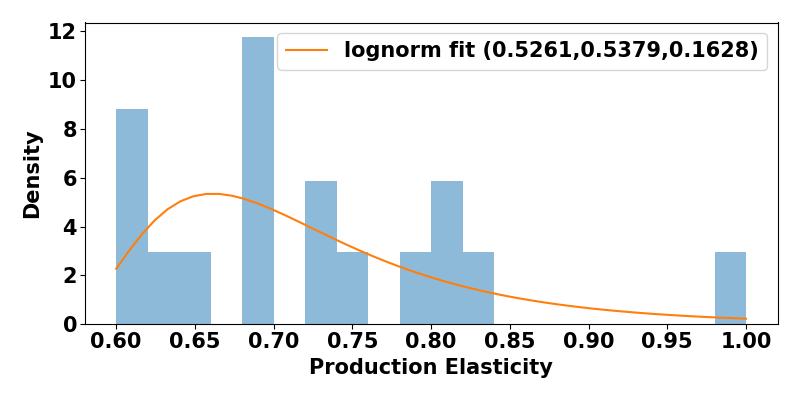}
    \caption{Histogram of fitted production elasticities for labor scaled to lie within $[0.6,1.0]$, used for verifying microeconomic stylized facts related to firms.}
    \label{fig:f_alpha_fit2}
\end{figure}

For verifying microeconomic stylized facts related to firms in Section \ref{subsubsec:firm_facts}, we first scale this set of elasticities to lie within $[0.6,1.0]$. Next, we determine the best-fit distribution for the resulting histogram using the distfit package \cite{Taskesen_distfit_is_a_2020}, identifying the lognormal distribution as the best fit, as shown in Figure \ref{fig:f_alpha_fit2}.
% \section{Household skill distribution}
% Setting exponential parameter so that $P(\omega>2)<0.05$.



% \section{Convergence of training rewards}

% \begin{figure}[h!]
%     \centering
%     \includegraphics[width=0.9\linewidth]{stylized_facts/diff_norm_reward.jpg}
%     \caption{Percentage difference in the moving average of training rewards as a function of training episodes for the first economic configuration.}
%     \label{fig:diff_rewards1}
% \end{figure}
% \begin{figure}[h!]
%     \centering
%     \includegraphics[width=0.9\linewidth]{utility/diff_norm_reward.jpg}
%     \caption{Percentage difference in the moving average of training rewards as a function of training episodes for the second economic configuration.}
%     \label{fig:diff_rewards2}
% \end{figure}

\end{appendices}


\bibliography{sn-bibliography}% common bib file

\end{document}